\documentclass{aa}
\usepackage{psfig}
\newcommand\AL{$A_{\rm Ly\alpha}$}
\newcommand\AMI{$A_{\rm{M/I}}$}

\newcommand\eg{{e.g.,~}}
\newcommand\etal{{et al.~}}
\newcommand\ie{{i.e.,~}}
\newcommand\Lya{Ly$\alpha$}
\newcommand\Ha{H$\alpha$}
\newcommand\Hbeta{H$\beta$}
\newcommand\NV{\hbox{N~$\rm V$}}
\newcommand\NVfull{\hbox{N~$\rm V$}~$\lambda$~1240}
\newcommand\CIVfull{\hbox{C~$\rm IV$}~$\lambda\lambda$~1549}
\newcommand\HeIIfull{\hbox{He~$\rm II$}~$\lambda$~1640}
\newcommand\CIIIfull{\hbox{C~$\rm III$]}~$\lambda$~1909}
\newcommand\CIIfull{\hbox{C~$\rm II$]}~$\lambda$~2326}
\newcommand\OIIfull{\hbox{[O~$\rm II$]}~$\lambda$~3727}
\newcommand\OIIIfull{\hbox{[O~$\rm III$]}~$\lambda$~5007}
\newcommand\MgIIfull{\hbox{Mg~$\rm II$}~$\lambda$~2800}
\newcommand\NeV{\hbox{[Ne~$\rm V$]}~$\lambda$~3426}
\newcommand\NeIII{\hbox{[Ne~$\rm III$]}~$\lambda$~3869}
\newcommand\CIV{\hbox{C~$\rm IV$}}
\newcommand\HeII{\hbox{He~$\rm II$}}
\newcommand\CIII{\hbox{C~$\rm III$]}}
\newcommand\CII{\hbox{C~$\rm II$]}}
\newcommand\MgII{\hbox{Mg~$\rm II$}}
\newcommand\OII{\hbox{[O~$\rm II$]}}
\newcommand\OIII{\hbox{[O~$\rm III$]}}
\newcommand\HI{\ion{H}{i}}

\newcommand\nodata{{...}}
\newcommand\araa{{ARA\&A}}
\newcommand\aap{{A\&A}}
\newcommand\aasup{{A\&AS}}
\newcommand\aj{{AJ}}
\newcommand\apj{{ApJ}}
\newcommand\apjl{{ApJ}}
\newcommand\apjs{{ApJS}}
\newcommand\mnras{{MNRAS}}
\newcommand\nature{{Nature}}
\newcommand\pasp{{PASP}}
\def\spose#1{\hbox to 0pt{#1\hss}}
\newcommand\simlt{\mathrel{\spose{\lower 3pt\hbox{$\mathchar"218$}}
     \raise 2.0pt\hbox{$\mathchar"13C$}}}
\newcommand\simgt{\mathrel{\spose{\lower 3pt\hbox{$\mathchar"218$}}
     \raise 2.0pt\hbox{$\mathchar"13E$}}}

\begin{document}

\thesaurus{(13.18.1;11.06.1;11.05.2)}
\title{A Statistical Study of Emission Lines from High Redshift Radio Galaxies \thanks{Table A.1 is also  available in electronic form at the CDS via anonymous ftp to cdsarc.u-strasbg.fr (130.79.128.5) or via http://cdsweb.u-strasbg.fr/Abstract.html, and Table B.1 is only available electronically from http://link.springer.de/link/service/journals/00230/index.htm}}
\titlerunning{Statistical study of emission lines from HzRGs}

\subtitle{}

\author{Carlos De Breuck \inst{1,2} \and Huub R{\"o}ttgering \inst{1} \and George Miley \inst{1} \and Wil van Breugel \inst{2} \and Philip Best \inst{1}}

\authorrunning{Carlos De Breuck \etal}


\institute{Sterrewacht Leiden, Postbus 9513, 2300 RA Leiden, The
Netherlands; debreuck,rottgeri,miley,pbest@strw.leidenuniv.nl \and Institute
of Geophysics and Planetary Physics, Lawrence Livermore National
Laboratory, L-413, Livermore, CA 94550, U.S.A.; wil@igpp.ucllnl.org
}
	
\date{Received 2000 May 12; accepted 2000 July 24}
	
\maketitle

\begin{abstract}   

We have compiled a sample of 165 radio galaxies from the literature to study the properties of the extended emission line regions and their interaction with the radio source over a large range of redshift $0<z<5.2$. For each source, we have collected radio (size, lobe distance ratio and power) and spectroscopic parameters (luminosity, line width and equivalent width) for the four brightest UV lines. We also introduce a parameter \AL\ measuring the asymmetry of the \Lya\ line, assuming the intrinsic redshift of the line is the same as that for the \HeIIfull\ line, and show that this parameter is a good measure of the amount of absorption in the \Lya\ line.

Using these 18 parameters, we examine the statistical significance of all 153 mutual correlations, and find the following significant correlations: (i) \Lya\ asymmetry \AL with radio size $D$ and redshift $z$, (ii) line luminosity with radio power, (iii) line luminosities of \Lya, \CIV, \HeII\ and \CIII\ with each other, and (iv) equivalent widths of \Lya, \CIV, \HeII\ and \CIII\ with each other.
We interpret the correlation between redshift and \AL\ as an increase in the amount of \HI\ around radio galaxies at $z>3$. The almost exclusive occurrence of \HI\ absorption in small radio sources could indicate a denser surrounding medium or an un-pressurized, low density region, as suggested by Binette \etal (2000)\nocite{bin00}. Correlations (ii) to (iv) provide evidence for a common energy source for the radio power and total emission line luminosity, as found in flux density-limited samples of radio sources.

The luminosity of the \Lya\ line relative to the other emission lines and the continuum shows a strong increase at $z \simgt 3$, coincident with the increase in the amount of associated \HI\ absorption. This indicates an increased abundance of hydrogen, both ionized and neutral, which may well be the reservoir of primordial hydrogen from which the galaxy is forming. This metallicity evolution is also seen in the nitrogen abundance, which shows a variation of more than an order of magnitude, with the $z>3$ radio galaxies occupying only the $Z<2Z_{\odot}$ region.

To examine the ionization mechanism of the extended emission line regions in HzRGs, we plot the UV emission line data in line-ratio diagnostic diagrams. The diagrams involving the high ionization \CIV, \HeII\ and \CIII\ lines seem to confirm previous results showing that AGN photo-ionization provides the best fit to the data. However, these models cannot fit the \CII/\CIII\ ratio, which lies closer to the predictions for the highest velocity shock ionization models. We note that the \CII\ line is five times more sensitive to shock ionization than the high ionization UV lines, and show that a combination of shock and photo-ionization provides a better overall fit to the integrated spectra of HzRGs. A substantial contribution from shock ionization will show up first in shock sensitive lines like \CII\ or \MgII.
We also confirm the findings of Best, R\"ottgering \& Longair (2000b)\nocite{bes00b} that shock ionization occurs almost exclusively in small radio sources, and show that the angular size distribution can indeed explain the differences in three HzRG composite spectra. Because most HzRGs have radio sizes $\simlt 150$~kpc, their integrated spectra might well contain a significant contribution from shock ionized emission.

\end{abstract}

%

\section{Introduction}

During the last two decades, high redshift radio galaxies (HzRGs) have been used as probes of galaxy formation. Out to $z\sim 1$, they are uniquely identified with massive ellipticals (\eg \cite{lil84}; \cite{bes98}). There are strong indications that this is also true at higher redshifts, mostly based on the remarkably strong correlation in the Hubble $K-z$ diagram out to $z=5.19$ (\cite{lil89}; \cite{eal97}; \cite{wvb98}; \cite{wvb99}). By studying radio galaxies over a large range of redshift, we can thus study the formation and evolution of massive galaxies.

The spectra of HzRGs are mostly dominated by characteristic extended emission lines, indicative of a halo of ionized gas. The most prominent line in these extended emission line regions is \Lya: the luminosity can reach $\sim 10^{45}$~erg~s$^{-1}$ and the spatial extent can be up to $\sim 150$~kpc (\eg\ \cite{oji96}, \cite{ada97}). This allows a detailed study of the kinematics of the gas, both close to the galaxy, where interactions with the radio jets can be studied, as well as at large distances from the AGN where the gas is still undisturbed and should trace the primordial distribution of the gas in these massive galaxies.
Within the extent of the radio source, the high line velocity widths ($\sim 1500$~km~s$^{-1}$) and disturbed morphologies (\eg \cite{vil99b}, \cite{bic00}) of the emission line gas indicate a strong interaction with the radio jets. This interaction is also obvious in statistical comparisons of the radio and optical morphologies in $z \simgt 0.6$ radio galaxies: (i) the UV and optical emission is often aligned with the radio source (\eg \cite{mcc87}, \cite{cha87}), (ii) the radio and emission-line gas morphological asymmetries are strongly correlated (\cite{mcc91a}), and (iii) the rest-frame $U-$ or $B-$band morphology depends on radio size (\cite{bes96}).

The bright \Lya\ emission profiles frequently show narrow absorption caused by neutral \HI\ in the HzRG (\cite{oji97}). This \HI\ absorption seems to occur almost exclusively in small radio sources, prompting van Ojik \etal\ to suggest that the \HI\ absorption indicates a dense intergalactic region which confines the radio source. However, the detection of highly ionized \CIVfull\ absorption in 0943$-$242 at $z=2.93$ suggests that the absorption is located further out in a low-metallicity gas shell (\cite{bin00}), providing us with a new tool to probe the outer regions of forming massive galaxies.

The radio power and emission line luminosities are found to be correlated (\eg \cite{mcc93}, \cite{wil99}), indicating that the central AGN is the common energy source for both. The most likely mechanisms for the transformation of this AGN energy into emission line luminosity are photo-ionization by an anisotropic UV source and shock excitation. Because the ionizing spectra of these mechanisms are quite different, they will lead to differences in the emission line spectra, which can be used to determine the dominating source of ionization. Such studies using UV line ratio diagrams suggest that the main ionizing mechanism in HzRGs is nuclear photo-ionization (\eg \cite{vil97}, \cite{all98}, hereafter VM97 and ADT98). However, a recent study of $z\sim 1$ radio galaxies finds that shock ionization is important in most of the smaller ($< 150$~kpc) sources (\cite{bes00b}, hereafter BRL00). Because the radio sources in HzRGs generally have sizes $< 150$~kpc, this is in apparent contradiction with the results of VM97 and ADT98 (unless there is a drastic change in the ionization mechanism between $z\sim 1$ and $z \simgt 2$).

From the above, it is clear that the properties of the gas in HzRGs can provide an important diagnostic to study various processes in forming massive galaxies, such as star formation, chemical enrichment of the interstellar matter, and the influence of the AGN through shocks or photo-ionization. Detailed observations of several representative individual galaxies (such as 4C~41.17, \cite{cha90}, \cite{dey97},\cite{bic00}) can be used to determine the relative importance of these mechanisms, but these should be complemented with a search for statistical relations between the emission line properties and other radio properties of a large sample of HzRGs. Such a study for $z \simlt 2.5$ radio galaxies (\cite{bau00}) finds systematically larger line widths and velocity field amplitudes at $z>0.6$ than at lower redshifts, but remains inconclusive on the origin: gravitational or due to jet-cloud interactions.

To allow such a statistical study over a continuous redshift range $z=0$ to $z=5.2$, we have compiled from the literature the emission line properties of a large sample of radio galaxies, together with relevant radio properties.  In this paper, we first describe the compilation of our HzRG sample (\S 2), and then go on to discuss the determination of the different radio and spectroscopic parameters (\S 3). In \S 4, we perform a statistical analysis of correlations between these parameters, taking the sometimes strong selection effects into account. In \S 5, we use diagnostic line-ratio diagrams to examine the ionization mechanisms in HzRG, and we discuss the implications of our results for the nature of HzRGs in \S 6. We present our conclusions in \S 7.

Throughout, we shall assume $H_0=50$~km~s$^{-1}$Mpc$^{-1}$, $q_0$=0.5, and $\Lambda=0$, unless otherwise stated, but our results do not depend on the adopted cosmology. We shall abbreviate the emission lines as follows: \NV\ for \NVfull, \CIV\ for \CIVfull, \HeII\ for \HeIIfull, \CIII\ for \CIIIfull, \CII\ for \CIIfull, \MgII\ for \MgIIfull, \OII\ for \OIIfull, and \OIII\ for \OIIIfull.


\section{Sample selection}

Table \ref{radiosurveys} lists nine samples designed to find HzRGs. The surveys can be divided into two classes: (i) several large flux density-limited surveys such as the 3CR (\cite{spi85}), BRL sample (\cite{bes99}), MRC (\cite{mcc96}), and 6C, 7C and 8C surveys (\eg \cite{lac99}); (ii) several ``filtered'' surveys, which have been designed to find the highest redshift objects. For the latter, the radio spectral index is most often used (ultra steep spectrum sources, \eg \cite{deb00a}), sometimes in combination with an angular size upper limit (\eg \cite{blu98}). Alternatively, the filter consists of a flux density interval centered around the peak in the source counts around 1~Jy (\eg \cite{ali82}).

\begin{figure}
\psfig{file=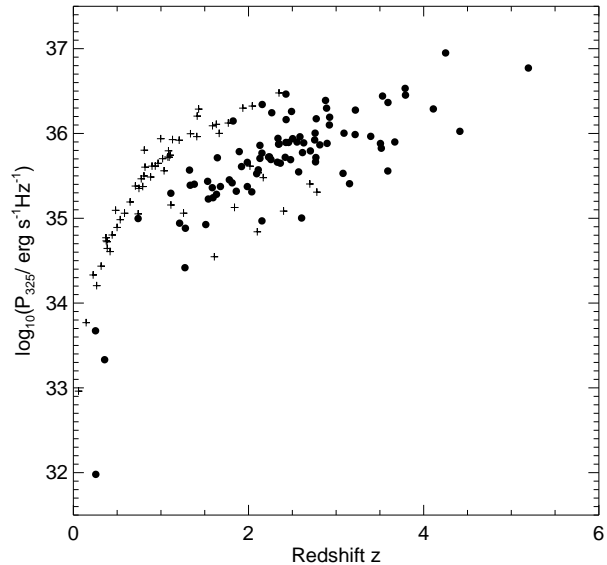,width=8.8cm}
\caption[]{Radio power at 325~MHz against redshift. Sources from un-filtered surveys are plotted with + signs, and sources from samples with spectral index and/or radio size filters are plotted as filled circles. Note the tight correlation at $z<1$ due to the flux density limited 3C survey, and the filtered surveys that fill up the lower power regions at $1<z<4$.}
\label{Pz}
\end{figure}

These two types of samples are complementary, in the sense that the complete surveys can provide information on the entire range of values of the parameters that were used as a high redshift filter (usually spectral index or radio size), while the filtered samples can extend the radio power and redshift coverage of the un-filtered samples. For example, the addition of data from filtered surveys partially compensates for the Malmquist bias in the flux density-limited surveys (see Fig. \ref{Pz}). We therefore compiled spectroscopic data on HzRGs from the samples listed in Table \ref{radiosurveys}, augmented with four sources from other small surveys. We shall concentrate on radio galaxies at $z>2$, because above this redshift, the brightest UV lines (\Lya, \CIV, \HeII, and \CIII) can be observed with ground based optical spectrographs.

\begin{figure}
\psfig{file=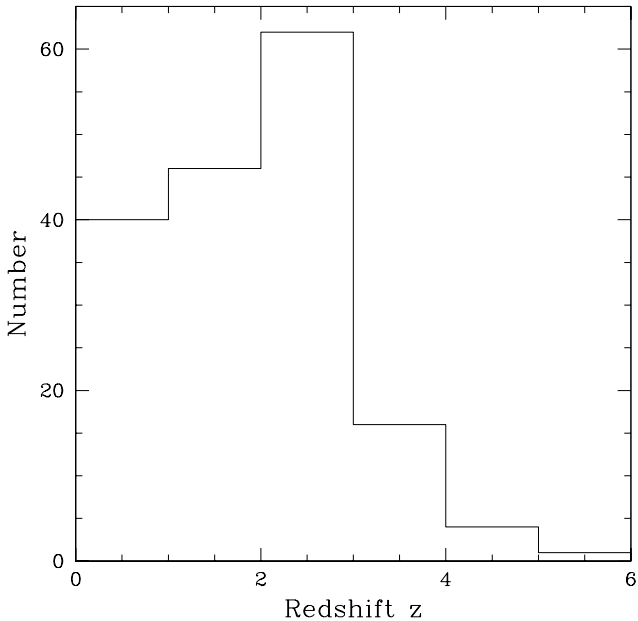,width=8.8cm}
\caption[]{Redshift distribution of the sample of $z>2$ radio galaxies with spectroscopic data,          and the lower redshift extension.}
\label{zhis}
\end{figure}

Of the 145 known $z>2$ radio galaxies, 78 have published spectroscopic parameters including line fluxes, equivalent widths, and line widths.
In order to fully investigate redshift dependence, we consider also spectroscopic data for the $z<2$ sources from the samples that provided sources at $z>2$.  Our final sample contains a similar number of sources at $z<2$ and at $z>2$ (Fig. \ref{zhis}). From Figure \ref{Pz}, we can see that we indeed include sources at $z \simgt 2$ with radio powers that are more than an order of magnitude lower than in the un-filtered samples. In appendix A, we list all 167 sources.
\begin{table*}
\caption[]{Radio surveys used to construct our sample of radio galaxies.}
\footnotesize
\begin{tabular}{lccccl}
\hline
Survey & Flux density limit & Spectral index limit & Angular size limit & $n_{spec}$ & Reference \\
\hline
3C     & $S_{178} > 10$~Jy & none & none & 15 & \cite{lai83} \\
BRL    & $S_{408} > 5$~Jy & none & none & 39 & \cite{bes99} \\
MRC    & $S_{408} > 0.95$~Jy & none & none & 19 & \cite{mcc96} \\
B2~1~Jy & $1 < S_{408} < 2$~Jy & none & none & 3 & \cite{ali82} \\
6C/7C/8C & varies$^a$ & none & none & 7 & \cite{lac99} \\
6C$^*$ & $0.96 < S_{151} < 2.00$~Jy & $\alpha_{151}^{4850} < -0.981$ & $\Theta < 15\arcsec$ & 2 & \cite{blu98}\\ 
WN/TN   & $S_{1400} > 10$~mJy & $\alpha_{325}^{1400} < -1.30$ & $\Theta < 1\arcmin$ & 34 & \cite{deb00a} \\
USS    & varies$^b$ & $\alpha < -1.0^b$ & none & 30 & \cite{rot94} \\
MG     & $S_{5000} > 50$~mJy & $\alpha_{1400}^{4800} < -0.75$ & $\Theta < 10\arcsec$ & 14 & \cite{ste99} \\
\hline
\end{tabular}

$^s$ The Cambridge surveys listed in Table~1 of Lacy \etal (1999)\nocite{lac99} consist of five sub-samples, each having different flux density limits.

$^b$ The USS sample of R\"ottgering \etal (1994)\nocite{rot94} consists of nine sub-samples, each with different flux density and spectral index limits. See Table 4 of R\"ottgering \etal (1994) for details.

\label{radiosurveys}
\end{table*}
%

%
%

\section{Determination of source parameters}
We now describe the radio and spectroscopic parameters from the literature papers and some derived quantities, as listed in Table A.1.

\subsection{Radio parameters}
\subsubsection{Radio power}
In order to obtain a uniform measurement of the radio flux densities, we determined the low frequency radio flux density from the 325~MHz WENSS ($\delta>+28$\degr; \cite{ren97}) or the 365~MHz Texas radio survey ($-35$\degr$<\delta<+28$\degr; \cite{dou96}), and the 1.4~GHz radio flux density from the NVSS (\cite{con98}). This procedure finds radio flux densities for 88\% of the sources in our HzRG sample. Of the 19 missing sources, one (USS~0529-549) lies outside the area covered by the surveys and one (VLA~J123642+621331) is too faint to be detected in any of the three surveys. The remaining sources have not been detected in the incomplete Texas catalogue.
Because no other deep large area radio surveys at frequencies below 325~MHz or above 1.4~GHz exist\footnote{The Cambridge 6C, 7C and 8C surveys only cover areas of the sky at $\delta>$30\degr, 20\degr\ and 60\degr, respectively (\cite{hal93},\cite{ril99},\cite{ree90}), while the Texas surveys covers $\delta>-35$\degr (\cite{dou96}).}, we could not correct for the spectral curvature when performing a K-correction in the calculation of the rest-frame mono-chromatic radio power. Because the radio spectra of HzRGs are predominantly concave, neglecting this spectral curvature will generally lead to an over-estimation of the radio power of the highest redshift sources. Neglecting a spectral curvature of $\Delta \alpha = 0.4$ between rest and observed frequency at $z=5$ will overestimate the radio power up to a factor of two. From the spectral curvature in a large USS sample (Figure 9 of De Breuck \etal 2000a\nocite{deb00a}), we expect only 30\% of the sources in our radio galaxy sample to have more spectral curvature, while all our sources are at lower redshifts. We therefore believe that our rest-frame radio powers are accurate to within a factor of two.

\subsubsection{Radio size}
To compile the data on the projected radio sizes, we generally took the largest separation between the components of the radio source from the same reference as the spectroscopy data (listed in appendix A). For the sources of R\"ottgering \etal\ (1997)\nocite{rot97}, we used the radio sizes from R\"ottgering \etal\ (1994)\nocite{rot94}, for the MRC/1~Jy sample of McCarthy \etal (1996) the sizes are from Kapahi \etal (1998)\nocite{kap98}, and for the MG sample of Stern \etal (1999)\nocite{ste99}, the sizes are from Lawrence \etal (1986) \nocite{law86}.

\subsubsection{Radio lobe distance ratio Q}
Carilli \etal (1997) and Pentericci et al. (2000) \nocite{car97} \nocite{pen00} have obtained 4.7~GHz and 8.2~GHz VLA observations of 64 HzRGs. For the objects for which they identified a radio core, we measured the radio lobe distance ratio Q from their 8.2~GHz contour plots\footnote{Q is defined by McCarthy, van Breugel \& Kapahi (1991) \nocite{mcc91a} as the ratio of the distance from the radio core to the more distant radio hotspot divided by the distance from the radio core to the closer of the radio hotspot.}. We also measured the Q values in the sample of Best, R\"ottgering \& Lehnert (1999)\nocite{bes99}, and took published values for 3CR galaxies from Best \etal (1995)\nocite{bes95} and McCarthy \etal\ (1991)\nocite{mcc91a}. For small and faint sources this parameter is difficult to determine, which limits the possibility to examine the dependency of Q, especially at the highest redshifts, where the radio size of the sources in our sample is smaller.

\subsubsection{Other radio parameters}
We considered to use the lobe flux density ratio R (\eg \cite{mcc91a}), which could provide a rough indication of orientation effects. However, too few published high resolution radio data are available at present to perform a statistically significant study of the correlations involving this parameter.

\subsection{Spectroscopic parameters}
\subsubsection{Line fluxes}
We used published line fluxes throughout this paper (see Appendix A for references). The apertures used to extract the one-dimensional spectra were generally chosen to include all the flux of the most spatially extended emission line. The spread in aperture sizes and depths of the spectra will give rise to an increased scatter when comparing line luminosities from different samples. Some objects were observed under non-photometric conditions, and so we only include these sources for line-ratio studies. We have also determined $5\sigma$ upper limits to the flux of \NV\ in 45 objects. Because line ratios are least sensitive to uncertainties in the relative flux calibration, we shall emphasise studies involving line ratios at the expense of studies involving detailed kinematics of the emission lines, which can only be determined from the brightest line, \ie\ \Lya.

\subsubsection{\Lya\ asymmetry}
The \Lya\ emission in HzRGs often shows absorption profiles caused by \HI\ surrounding the radio galaxy, (\eg \cite{oji97}, \cite{rot95}, \cite{dey99}, \cite{deb99}). This absorption preferentially occurs on the blue side of the emission line, which leads to a characteristic triangular shape of the \Lya\ line in two dimensional spectra. Such asymmetries in \Lya\ are often also observed in other objects with strong \Lya\ emission at very high redshift (\eg \cite{dey98}), and might also have a contribution from intervening \HI\ absorbers along the cosmological line of sight.
\begin{figure*}
\psfig{file=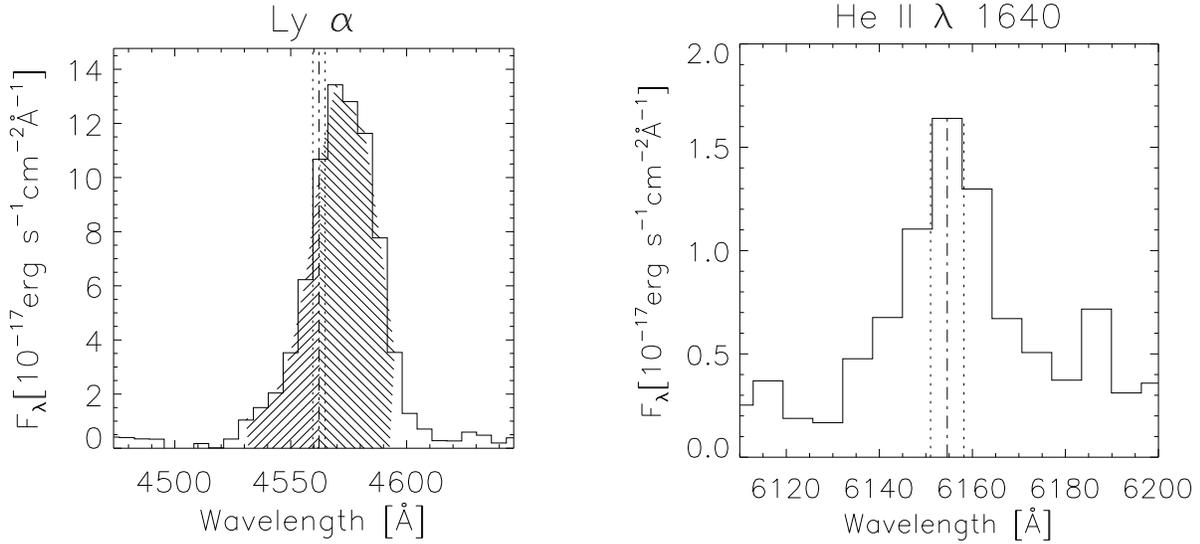,width=17cm}
\caption[]{Example of the determination of the \Lya\ asymmetry parameter \AL. The right panel shows the position of the \HeII\ line (dash-dotted line) determined by a Gaussian fit with the uncertainties indicated by dotted lines. The left panel shows the position of the peak of \Lya\ as predicted from the \HeII\ fit, with the errors indicated. The left and right hatched areas indicate the 2000~km/s intervals used to calculate the flux on both sides of the assumed systemic redshift. The \AL\ derived from this spectrum is \AL = $-0.49$, with the range of uncertainties going from $-0.58$ to $-0.39$.}
\label{tn0920}
\end{figure*}
%

While high resolution spectroscopy of these \Lya\ lines allows one to determine characteristics of the \HI\ absorber (\cite{rot95}, \cite{oji97}), this is rarely possible with their discovery spectra, which have typical resolutions of a few hundred. To derive some information on the \HI\ absorption from these low-resolution spectra, we therefore introduce a parameter which measures the relative flux on the blue and red sides of the assumed systemic redshift. Because absorption near the maximum of the \Lya\ emission can shift the observed peak of the profile by as much as 500~km~s$^{-1}$ (\cite{rot97}), we have to use other emission lines to determine the systemic redshift. The only other strong lines in HzRGs observed simultaneously with \Lya\ are \CIV, \HeII, and \CIII. From these, \HeII\ is the most appropriate line to use, because unlike \CIV, \HeII\ is not a resonant line, so the profile should be less affected by absorption.

We obtained one-dimensional spectra from the samples of R\"ottgering \etal (1997)\nocite{rot97}, Stern \etal (1998) \nocite{ste99}, and De Breuck \etal (2000b) \nocite{deb00b}. In 31 objects, we were able to obtain a good Gaussian fit of the \HeII\ line. We approximated the error in the position of the peak of the \Lya\ line as the quadratic sum of the error in the \HeII\ Gaussian fit and the error in the wavelength calibration, the latter being approximated as a quarter of the dispersion of appropriate spectrum (the resolution of the different spectra varies from $\sim$10\AA\ to $\sim$25\AA).

\begin{figure}
\psfig{file=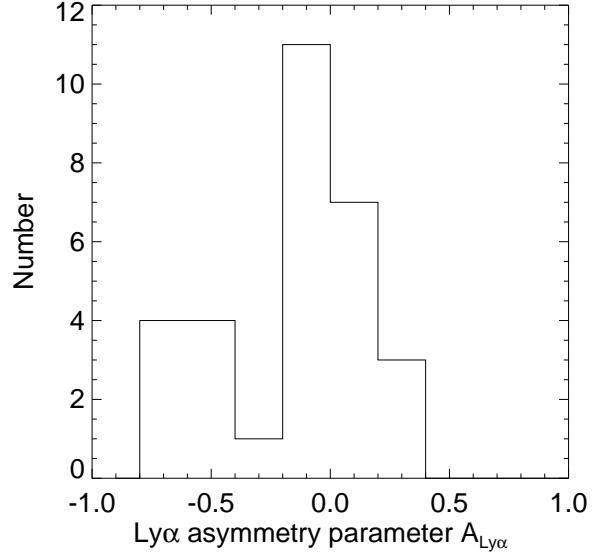,width=8.8cm}
\caption[]{Distribution of the \Lya\ asymmetry parameter \AL. Note the preference for blue-ward absorption (negative \AL\ values).}
\label{ALhistogram}
\end{figure}

We define the \Lya\ asymmetry parameter $$A_{\rm Ly\alpha}=\frac{F_{blue}-F_{red}}{F_{blue}+F_{red}},$$ where $F_{blue}$ is the flux within 2000~km~s$^{-1}$ blue-ward of the systemic \Lya\ wavelength and $F_{red}$ is the flux within 2000~km~s$^{-1}$ red-ward of the systemic \Lya\ wavelength (see Fig. \ref{tn0920} for an example). Because in the lowest resolution spectra, these 2000~km~s$^{-1}$ intervals often include only a few dispersion elements (pixels) of the spectrogram, we include only the percentage of the flux in the lowest and highest bin that lies within the 2000~km~s$^{-1}$ interval. A value of \AL=$-1$ means total absorption of the flux on the blue side, a completely symmetrical profile has \AL=0, and positive \AL\ values indicate absorption on the red side. We calculated the range of allowed \AL\ values from the error in the predicted systemic wavelength of \Lya. Because the peak sometimes falls on a steep side of the observed \Lya\ profile, this can lead to asymmetric errors in \AL. We excluded three sources where the range in \AL was larger than 1.5 due to their large uncertainties. We could accurately determine the \AL\ parameter for 25 HzRGs. The distribution (Fig. \ref{ALhistogram}) shows a strong peak around \AL=0 (no asymmetry) and a secondary peak around \AL=$-$0.6 (blue absorption), but no sources with strong red-ward absorption. This is consistent with the trend seen from high resolution spectroscopy, where the median velocity of the \HI\ absorbers is blue-shifted by 100~km~s$^{-1}$ with respect to the peak emission redshift (\cite{oji97}).
%

\begin{figure}
\psfig{file=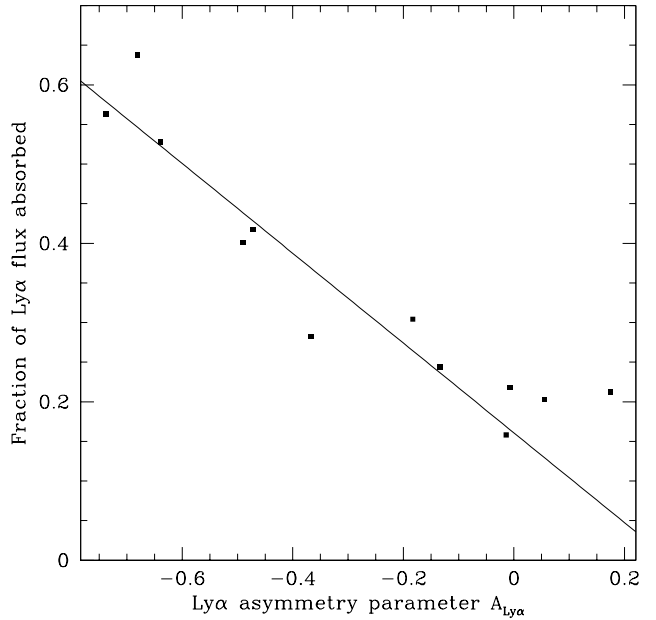,width=8.8cm}
\caption[]{The fraction of \Lya\ flux absorbed plotted against the \Lya\ asymmetry parameter (see text for details). Note than no more than 65\% gets absorbed, and that the asymmetry parameter is a good measure of the total amount of absorption, as indicated by the linear fit.}
\label{asymfluxdef}
\end{figure}

For 12 HzRGs, the resolution and signal to noise of the \Lya\ profile was sufficient to extrapolate over the absorption profiles while fitting the emission with a single Gaussian profile (see \cite{deb99} for an example). In cases where the profile is clearly non-Gaussian, we also included a Voigt function. We again used the \HeII\ redshift to predict the peak position of the Gaussian \Lya\ emission (in two cases a variation of 1--2\AA\ provided a better fit, which is consistent with the wavelength calibration errors). We varied the peak flux and FWHM of the Gaussian until we obtained a good fit on one side of the profile. We use these crude Gaussian models to determine how much flux was emitted before the \HI\ absorption (integrating out to three times the FWHM of the Gaussian fit). We used this total flux  to calculate an approximate fraction of the \Lya\ flux that is absorbed. Figure \ref{asymfluxdef} compares this fraction with \AL. As expected, we find a strong correlation indicating that (i) our \AL\ parameter is a good approximation of the absorbed flux and (ii) that even in the most asymmetric profiles, the observed \Lya\ flux is only diminished by $\sim 50$\%, although we cannot exclude the possibility that sources with higher absorption fractions are missing from the samples because they are too weak to be detected. A limitation of our method is that it is insensitive to absorption which is centered near the peak of the emission, but such cases are rather rare (\cite{oji97}).

For two of the highest redshift HzRGs where \HeII\ falls outside the optical window, we use the fit in Figure \ref{asymfluxdef} to calculate the approximate \AL\ value. This is particularly important if we want to study the redshift evolution of \AL.
As a final note, it should be stressed that this procedure provides only a rough measure of the \Lya\ kinematics. The main goal of the \AL\ parameter is to examine the existence of strong trends with redshift, or parameters of the radio source. A detailed study of the kinematics is impossible with these low-resolution data, and is beyond the scope of this paper.

%

\section{Correlations between parameters}

\subsection{Selection effects}
The main problems we encounter when searching for correlations between the different parameters in our sample of HzRGs are the selection effects resulting from the search techniques used to find them and technical limitations of the spectrographs. A well known example of the first effect is the Malmquist bias in flux density limited samples. As discussed in \S 2, the addition of several filtered surveys with lower flux density limits alleviated this bias, especially at $2<z<4$, the range we shall concentrate on (Fig. \ref{Pz}). However, comparisons with $z<2$ radio galaxies will still be strongly affected by Malmquist bias.

More subtle selection effects are due to the wavelength-dependent sensitivity of the optical spectrographs. Weak emission line objects could well have been escaped detection at certain redshifts where the brightest lines are not observable. For example, fewer sources are known in the $1.5 \simlt z \simlt 2$ ``redshift desert'' where \OIIfull\ has shifted out of the optical window and \Lya\ has not yet entered, and the redshift will have to be based on weaker lines like \CIV, \HeII\, \CIII\ or \CII\ (\eg \cite{ste99}). At very high redshift ($z>3$), weaker lines red-ward of \Lya\ will shift to wavelengths where measurement is difficult due to fringing of the CCD and bright OH sky-lines (\eg \cite{ost92}). This should have only a limited effect on the line ratio diagnostics, because the spectra of $z>3$ radio galaxies are generally of higher quality than those at $2<z<3$: the highest redshift spectra were usually obtained with larger aperture telescopes, or the integration time was extended to search for confirming lines.

\subsection{Statistical tests}
Most previous studies have each concentrated on specific correlations between a few specified parameters in HzRGs. Here, we shall examine correlations between all of our measured parameters. Some apparent correlations between two parameters might be due to the correlation of one or both of these parameters with a third parameter (either by selection effects or by a real correlation). As argued by Macklin (1982)\nocite{mac82}, the statistical test that de-couples such dependent correlations, while retaining as much information as possible, is the Spearman partial rank correlation coefficient. We shall apply this test to determine if some correlations are not due to selection effects in one or both of the parameters.

In the first approximation, we simply exclude lower or upper limits from this analysis. For most parameters, this will not have a serious effect on the results, because less than 5\% of the data are lower or upper limits. However, 12\% of the linear sizes are upper limits, and an even higher percentage of the equivalent widths are lower limits (from 11\% for $W^{rest}_{\rm{HeII}}$ to 35\% for $W^{rest}_{\rm{Ly}\alpha}$). To treat these limits, or ``censored data'', in a statistically meaningful manner with minimal loss of information, a class of statistical methods called ``survival analysis'' has been developed (\eg \cite{iso86}). To test the correlations involving radio size and equivalent widths with survival analysis methods, we shall use the ASURV software package (\cite{iso90}, \cite{lav92}), which is incorporated into the NOAO reduction package IRAF. Three different correlation tests within ASURV are of importance to this work: Cox's hazard model, generalized Kendall's $\tau$, and generalized Spearman's $\rho$. We shall take the variables as correlated if all three of these models give significance levels\footnote{Note that for consistency with previous statistical work on HzRGs, by significance level we mean the significance level of the correlation, and not the probability of the null hypothesis (no correlation) being true, which is generally used in survival analysis.} $>$95\%. One of the deficiencies of survival analysis for astronomical purposes is that upper limits are assumed to be precisely known. Especially the lower limits to the equivalent widths are very uncertain due to the extreme faintness of the continuum emission. Survival analysis is not as effective in handling dependent correlations in the way the Spearman partial rank analysis can. We shall therefore only use survival analysis to check whether the addition of censored data changes the significance of a correlation which was found from our previous analysis.

\subsection{Spearman rank analysis}
To investigate every possible dependent correlation, we first calculated the Spearman rank correlation coefficient and the associated significance levels for all 153 possible combinations of source parameters in our HzRG sample. The results are presented in appendix B. The correlation coefficients and number of available parameters are tabulated in the lower left half of the Table, and the significance levels in the upper right half. We also calculated these values for a $H_0=65$~km~s$^{-1}$Mpc$^{-1}$, $q_0=0.15$ cosmology, but this changed the values of the correlation coefficients by less than than 0.1.

We find 33 correlations with significance levels $>$99\%, and 10 additional which are between 95\% and 99\% significant.
These correlations are:

\noindent$\bullet$ Redshift $z$ with radio size $D$, radio power $P_{325}$, radio spectral index $\alpha$ and line luminosity $L_{\rm{CIV}}$, $L_{\rm{HeII}}$ or $L_{\rm{CIII]}}$.

\noindent$\bullet$ Radio spectral index $\alpha$ with radio size $D$.

\noindent$\bullet$ Radio spectral index $\alpha$ with radio power $P_{325}$.

\noindent$\bullet$ Radio spectral index $\alpha$ with line luminosity (increasingly stronger correlation for weaker emission lines).

\noindent$\bullet$ Radio spectral index $\alpha$ with equivalent width of \CIV, \HeII\, and \CIII.

\noindent$\bullet$ Radio size $D$ with \Lya\ asymmetry \AL, and $L_{\rm{HeII}}$.

\noindent$\bullet$ Radio lobe distance ratio $Q$ with equivalent width $W^{rest}_{\rm{Ly}\alpha}$ and $W^{rest}_{\rm{HeII}}$.

\noindent$\bullet$ Radio power $P_{325}$ and $P_{1400}$ with all four line luminosities.

\noindent$\bullet$ Line luminosities of \Lya, \CIV, \HeII\ and \CIII\ with each other.

\noindent$\bullet$ \Lya\ luminosity with both \Lya\ equivalent width $W^{rest}_{\rm{Ly}\alpha}$ and \Lya\ line width $\Delta v_{\rm{Ly}\alpha}$.

\noindent$\bullet$ Equivalent widths of \Lya, \CIV, \HeII\ and \CIII\ with each other.

\noindent$\bullet$ Equivalent widths of fainter lines (\HeII\ and \CIII) with line widths of \Lya\ and to a lesser extent fainter lines.

We shall now examine if some of these correlation are due to selection effects, dependent correlations with other parameters, or are removed if sources with upper/lower limits are included. 

\subsection{Correlations influenced or caused by selection effects}
\subsubsection{Radio size and power vs. redshift}
The dependence of radio size $D$ on redshift $z$ and radio power $P_{325}$ has been examined by a number of authors (\eg \cite{nee95}, \cite{blu99a}). The range of often contradictory results is due to several selection effects (\eg in selection frequency) in the samples used to examine this correlation.   
Our sample of radio galaxies is too inhomogeneous and incomplete to address this question. For example the Malmquist bias in radio power (Fig. \ref{Pz}) will dominate the $P_{325} - z$ correlation. For $z>2$, the probability that $z$ and $P_{325}$ are uncorrelated drops from $1.15 \times 10^{-21}$ to $1.6 \times 10^{-4}$, indicating the Malmquist bias is strongly decreased, but not removed (see Fig. \ref{Pz}).
While part of the correlation might be due to an underlying real correlation (\eg between $z$ and $D$), we consider the effect of the selection effects too large and will not examine the significance of these relations.

\subsubsection{Radio spectral index vs. radio size and radio power}
In some samples designed to find HzRGs, the radio size $D$ has been used as an additional ``high redshift filter'' in combination with a spectral index cutoff. This will lead to selection effects in the redshift-spectral index relation, masking a possible real correlation.

The relation between spectral index and radio power is not independent, because we used the spectral index to calculate the rest-frame radio power of our sources.
%
\begin{figure}
\psfig{file=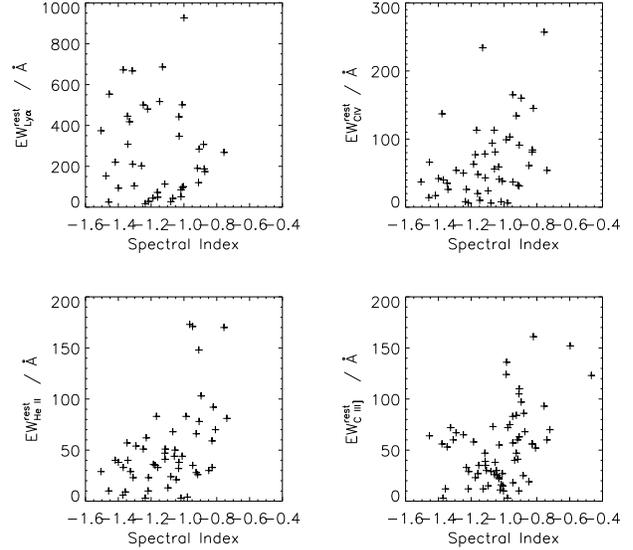,width=8.5cm}
\caption[]{Rest-frame equivalent width plotted against radio spectral index. Note that the trend for redder lines (\HeII\ and \CIII) is more offset towards flatter spectral indices than for the blue lines (\Lya\ and \CIV). This arises because at higher redshift (\ie steeper spectral index) the bluest lines either shift into less sensitive parts of the observable window, where they can only be detected if they have high equivalent widths.}
\label{spixEW}
\end{figure}
%

\subsubsection{Radio spectral index vs. line luminosity and equivalent width}
The luminosities of the UV emission lines and rest-frame equivalent widths appear to be correlated, with the weakest line (\CIII) showing the strongest correlation. However, due to the way we constructed our sample of radio galaxies, the sources with the flattest spectral indices are all at relatively low redshift ($z \simlt 2$). Because the line luminosities are strongly correlated with redshift (see next section), we only find the low-luminosity lines in the low-redshift flattest spectrum sources of our sample. 

The apparent correlation between radio spectral index and rest-frame equivalent width is also influenced by subtle redshift selection effects: the increasingly bluer lines will only be detected at higher redshifts due to the fixed wavelength range of the optical spectrographs. At the lowest redshifts, these lines will be in the bluest, least sensitive part of the CCD, and they will only be detected when they have a high equivalent width. This effect is obvious in Fig. \ref{spixEW}: the apparent correlation between rest-frame equivalent width and spectral index seems to shift towards flatter spectral indices in the redder lines, which are only observed in the lower redshift sources which have flatter radio spectral indices.

\subsubsection{Emission line luminosity vs. redshift}
\begin{figure}
\psfig{file=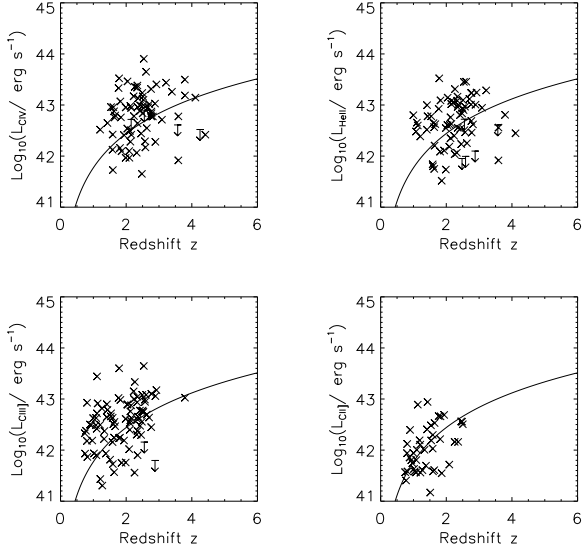,width=8.5cm}
\caption[]{Emission line luminosity for \CIV, \HeII, \CIII\, and \CII\ plotted against redshift. The plotted line indicates the luminosity as a function of redshift at which a line with an emission line flux of $1.0 \times 10^{-16}$erg~s$^{-1}$~cm$^{-2}$\AA$^{-1}$, near the detectability of 3-4m class telescopes, would be detected. Note that the higher wavelength lines are increasingly weaker, and lie closer to the detection limit. The lack of sources below this limit leads to an artificial redshift dependence of the line luminosity.}
\label{zlinelum}
\end{figure}

\begin{figure}
\psfig{file=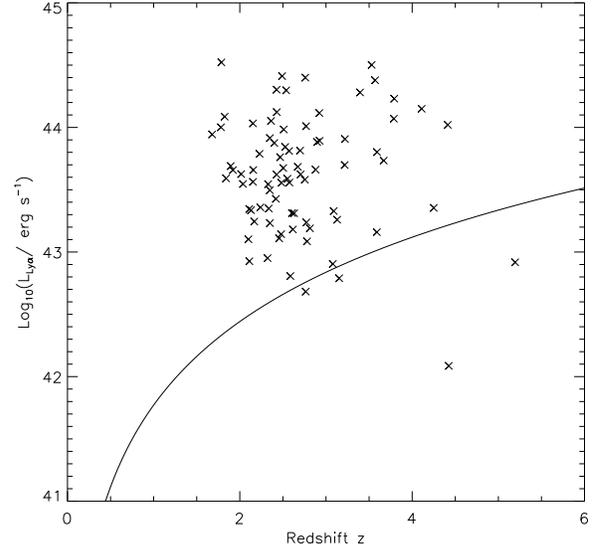,width=8.5cm}
\caption[]{Emission line luminosity for \Lya\ plotted against redshift. The line indicates a line with an emission line flux of $1.0 \times 10^{-16}$erg~s$^{-1}$~cm$^{-2}$\AA$^{-1}$, near the detectability of 3-4m class telescopes. Note that there is no artificial redshift dependence, as for the lines in Figure \ref{zlinelum}.}
\label{Lyalum}
\end{figure}

The apparent correlation between line luminosity and redshift $z$ is a less known selection effect. The correlation is weak or absent for \Lya, but increasingly stronger for \CIV, \HeII, and \CIII. The reason for this is obvious in Figures \ref{zlinelum} and \ref{Lyalum}, where we plot the emission line luminosities against redshift together with a curve which denotes the luminosity corresponding to an emission line with a flux of $1.0 \times 10^{-16}$erg~s$^{-1}$~cm$^{-2}$\AA$^{-1}$, which is close to the minimum detectable level of most of the observations. We clearly find a line-luminosity redshift degeneracy, as found for the radio powers. This degeneracy becomes more pronounced as the lines become weaker. The \Lya\ line is generally an order of magnitude brighter than the other UV lines. In addition, the two highest redshift radio galaxies have very weak \Lya\ (TN~J0924$-$2201 at $z=5.19$ and VLA~J123642+621331 at $z=4.42$; \cite{wvb99}, \cite{wad99}), and have been detected at flux levels an order of magnitude fainter than what has been attempted for the other lines. The lines red-ward of \Lya\ are increasingly fainter, and lie much closer to the detection limit, and will therefore more easily be missed.
%
%
\begin{table}
\caption[]{Results of survival analysis on the correlations between line luminosity and equivalent width. The significance indicates the probability a correlation is detected.}
\tiny
\begin{tabular}{cccccc}
\hline
\multicolumn{2}{c}{Variable} & Percentage & \multicolumn{3}{c}{Significance} \\
Independent & Dependent & Censored & Cox & Kendall & Spearman \\
\hline
$L_{\rm{Ly}\alpha}$ & $W^{rest}_{\rm{Ly}\alpha}$ & 36\% & 34.71\% & 85.00\% & 81.36\% \\
$L_{\rm{CIV}}$ & $W^{rest}_{\rm{CIV}}$ & 18\% & 99.92\% & 99.78\% & 99.65\% \\
$L_{\rm{HeII}}$ & $W^{rest}_{\rm{HeII}}$ & 11\% & 99.51\% & 99.25\% & 99.31\% \\
$L_{\rm{CIII]}}$ & $W^{rest}_{\rm{CIII]}}$ & 10\% & 70.19\% & 98.29\% & 97.85\% \\
\hline
\end{tabular}
\label{EWlinelumsurvival}
\end{table}

\subsubsection{Emission line luminosity vs. equivalent width}
The correlations between the equivalent width and line luminosities are also dominated by selection effects, namely the difficulty to detect objects with weak emission line fluxes and high equivalent widths. These sources probably do exist, as the objects for which there are lower limits to the equivalent widths frequently have low line luminosities. When we apply survival analysis models to the complete dataset (Table \ref{EWlinelumsurvival}), we find that the correlations with \Lya\ and \CIII\ are no longer significant. Given that both the detections and lower limits to the equivalent widths in these weak lines are highly uncertain, we interpret these correlations as too uncertain to be trustworthy.

\subsubsection{\Lya\ line width vs. equivalent width of \HeII\ and \CIII}
A final correlation which we consider dominated by selection effects is the correlation of equivalent widths of fainter lines (\HeII\ and \CIII) with line widths of \Lya. When the two sources with $\Delta v_{\rm{Ly}\alpha} > 3000$~km~s$^{-1}$ are not considered, the significance of these correlations drops to $<95$\%. Such sources probably have a broad component detected in \Lya, but not in the weaker lines.
\begin{figure}
\psfig{file=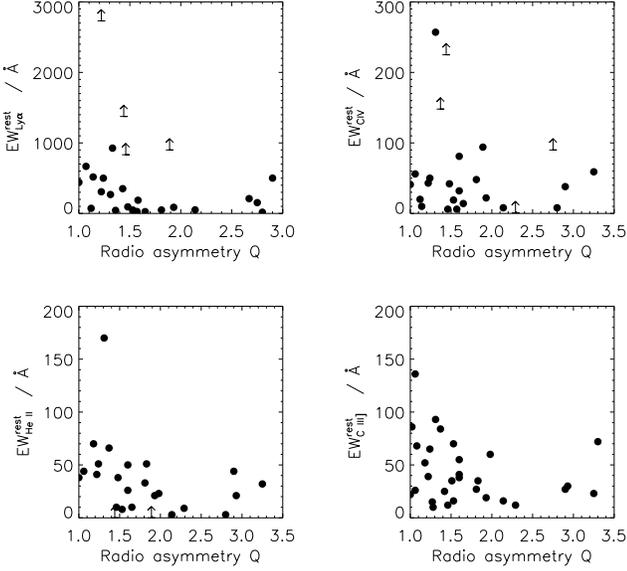,width=8.8cm}
\caption[]{Equivalent widths of the strong UV lines plotted against radio lobe distance ratio $Q$.}
\label{QEW}
\end{figure}
%

\subsection{Possible correlations}
A few parameters, for which limited data exist appear to show correlations. One set of such correlations are those between the radio lobe distance ratio $Q$ and rest-frame equivalent widths of the UV lines. From the plots of these four correlations (Fig. \ref{QEW}), we find that there is a dearth of highly asymmetric radio sources with large equivalent widths. 

One possible explanation of this effect can be found in the observing techniques used to determine the redshifts of HzRGs. In some objects where the correct identification of the host galaxy is uncertain, the spectroscopic slit will be positioned such as to increase the chances to detect strong line emission which can be used to determine the redshift. Most often, this means that the slit is aligned with the radio emission. In relatively symmetrical sources with a slight bending of the radio lobes, this could mean that the slit will not be perfectly centered on the host galaxy. In highly asymmetric sources, the slit will be more often centered near the presumed optical identification, which will increase the chance that host galaxy continuum emission is included, and lead to a lower equivalent width. Based on only $\sim$20 sources, we therefore consider this correlation suggestive. A sample of radio galaxies observed in a consistent way would be needed to fully examine the significance of this correlation.

\subsection{Probable correlations}
\subsubsection{Line luminosities and equivalent widths}
We find that the line luminosities of the four bright UV lines are all strongly correlated with each other. We also find a similar result for the rest-frame equivalent widths. This indicates that the different emission lines in HzRGs are powered by the same mechanism. We shall discuss this further in \S 5.

\subsubsection{Radio power vs. emission line luminosity}
\begin{figure}
\psfig{file=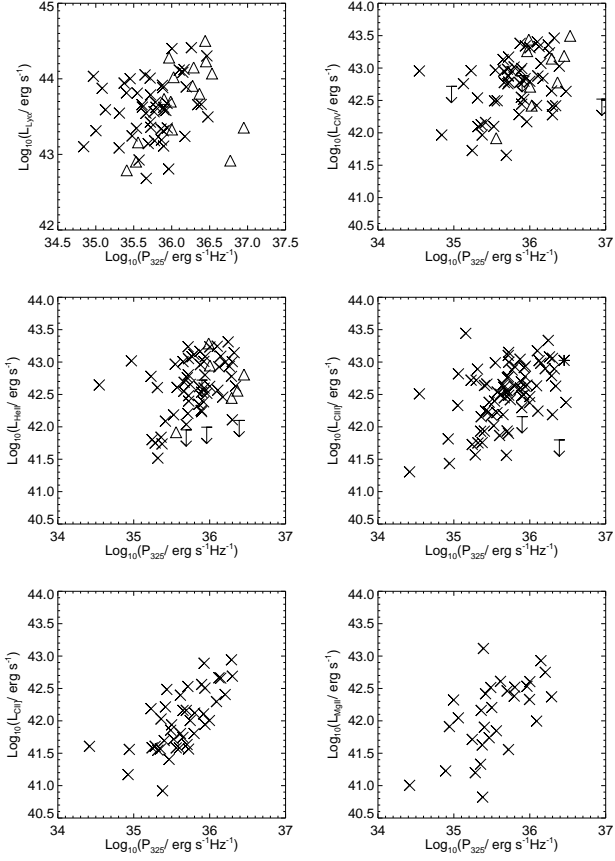,width=8.5cm}
\caption[]{Line luminosity for \Lya, \CIV, \HeII, \CIII\, \CII\ and \MgII\ plotted against radio power at 325~MHz. Crosses represent sources at $z<3$, triangles those at $z>3$.}
\label{radiopowerlinelum}
\end{figure}

To examine the correlation between radio power and emission line luminosity, we use the Spearman partial rank coefficient, including redshift as the third parameter, and thus taking the Malmquist bias of both parameters into account.
Table \ref{radiolineranks} presents the results of this analysis. Figure \ref{radiopowerlinelum} shows graphical representations of the correlations.

\begin{table}
\scriptsize
\begin{tabular}{lcccc}
\hline
Variables & Number & $r_{PL}$ & $r_{PL,z}$ & $\sigma$ \\
\hline
$P_{325}, L_{\rm{Ly}\alpha}$  & 67 & 0.35 (0.38) & 0.37 (0.38) & 3.06 (3.14) \\
$P_{325}, L_{\rm{CIV}}$       & 64 & 0.40 (0.28) & 0.29 (0.28) & 2.40 (2.27) \\
$P_{325}, L_{\rm{HeII}}$      & 61 & 0.35 (0.24) & 0.25 (0.24) & 1.94 (1.91) \\
$P_{325}, L_{\rm{CIII]}}$     & 79 & 0.47 (0.30) & 0.31 (0.30) & 2.78 (2.68) \\
$P_{325}, L_{\rm{CII]}}$      & 42 & 0.68 (0.73) & 0.70 (0.70) & 5.29 (5.39) \\
$P_{325}, L_{\rm{MgII}}$      & 30 & 0.62 (0.65) & 0.56 (0.56) & 3.25 (3.26) \\
\hline
\end{tabular}
\caption[]{Spearman partial rank correlation analysis for the correlations between radio power $P_{325}$, line luminosity of the bright UV lines and redshift. $r_{PL}$ is the Spearman rank coefficient of the radio power and line luminosity, and $r_{PL,z}$ is the Spearman partial rank coefficient, taking the selection effects with redshift into account. $\sigma$ is the significance of the partial rank correlation, which is equivalent to the deviation from a unit variance normal distribution if no correlation is present. Values in brackets are for a $H_0=65$~km~s$^{-1}$Mpc$^{-1}$, $q_0$=0.15 cosmology.}
\label{radiolineranks}
\end{table}
%

%

Table \ref{radiolineranks} shows that the selection effects with redshift have only a minor influence on the correlation between the radio powers and \Lya\, \CII\ and \MgII\ luminosity, but do artificially strengthen the correlations with \CIV, \HeII, and \CIII. This is because the radio power - redshift degeneracy becomes less pronounced, or absent at $z>2$, where the UV lines shift into the optical spectroscopy window, and because of the increasingly more important redshift selection effects with weaker emission lines, as discussed in \S4.3. We have examined this correlation using the radio power at 325~MHz and 1.4~GHz, and found only minir differences, with the $P_{325}$ yielding slightly stronger correlations. We prefer to use the $P_{325}$ because the higher frequency might have a larger contribution from a Doppler boosted core (see \eg \cite{blu99a}).

We find there is a $\simgt 2\sigma$ probability that the correlations between radio power and line luminosity do not arise from the correlations of individual parameters with redshift. We shall discuss this further in \S 6.2.

\subsubsection{\Lya\ asymmetry vs. Radio size, and vs. redshift}
The Spearman rank analysis (appendix B) indicates that there is an anti-correlation between radio size and \Lya\ asymmetry \AL\ with a confidence level of 99.86\%. We perform a Spearman partial rank analysis, because of the possibility of this arising artificially from the correlation between redshift and radio size D.
\begin{table}
\scriptsize
\begin{tabular}{lcrrr}
\hline
$x,y$ & Number & $r_{xy}$ & $r_{xy,z}$ & $\sigma$ \\
\hline
$D$, \AL & 25 & $-0.60 (     -0.59)$ & $-0.64 (      -0.64)$ & 3.48 (3.44) \\
\AL, $z$ & 25 & $ 0.33 (\;\;\;0.33)$ & $ 0.41 (\;\;\; 0.09)$ & 2.02 (2.06) \\
\hline
\end{tabular}
\caption[]{Spearman partial rank correlation analysis for the correlations between radio size $D$, \Lya\ asymmetry \AL, and redshift z. $r_{xy}$ is the Spearman rank coefficient, and $r_{xy,z}$ is the the Spearman partial rank coefficient, taking the selection effects with the third parameter into account. $\sigma$ is the significance of the partial rank correlation, as in Table \ref{EWlinelumsurvival}. Values in brackets are for a $H_0=65$~km~s$^{-1}$Mpc$^{-1}$, $q_0$=0.15 cosmology.}
\label{Asymcorrelations}
\end{table}
%

\begin{figure}
\psfig{file=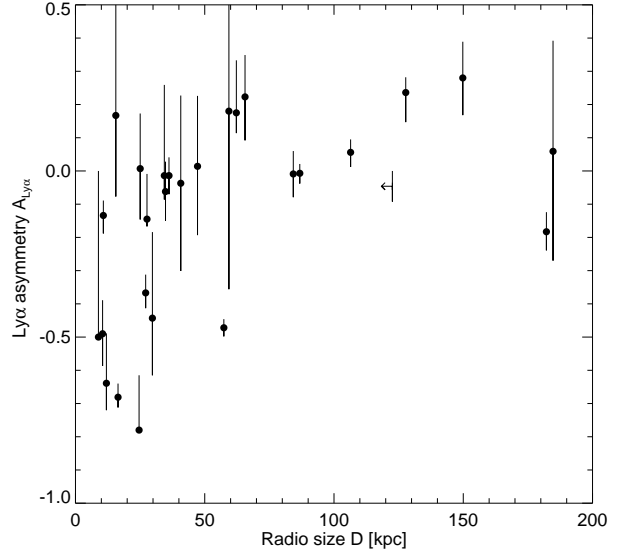,width=8.5cm}
\caption[]{\Lya\ asymmetry parameter \AL\ plotted against radio size. Error bars reflect the uncertainty in the wavelength calibration and fitting of the \HeII\ line, and need not be symmetrical. Note that strong blue \Lya\ absorption only occurs in smaller radio sources.}
\label{Dasym}
\end{figure}

From this analysis (Table \ref{Asymcorrelations}), we find that \AL\ appears to be correlated with radio size and redshift independently.
From Figure \ref{Dasym}, we see that all seven HzRGs with \AL$>0.3$ have radio sizes $D<60$~kpc. This confirms the results of van Ojik \etal (1997)\nocite{oji97}, who found from high resolution spectroscopy that \HI\ absorption preferentially occurs in radio galaxies smaller than 50~kpc. 
%

The correlation with redshift is less significant, but we lack sufficient data at $z>3$, because our method requires the detection of the \HeII\ line, which starts shifting out of the optical band at these redshifts. However, Dey (1999)\nocite{dey99} shows \Lya\ velocity profiles of four $z>3.5$ radio galaxies with the redshift predicted from \HeII\ or \CIII\ indicated, which all show strong blueward absorption. His results are not only consistent with the \AL$-D$ trend, but also show stronger absorption for the $z>4$ radio galaxies, as is also seen in TN~J1338$-$1942 at $z=4.11$ (\cite{deb99}). We discuss this result further in \S6.4.

\subsection{Summary of correlation analysis}
After consideration of all selection effects, we find 22 real correlations, viz.

\noindent$\bullet$ Radio size $D$ and redshift $z$ with \Lya\ asymmetry \AL.

\noindent$\bullet$ UV line luminosity with radio power (8 correlations).

\noindent$\bullet$ Line luminosities of \Lya, \CIV, \HeII\ and \CIII\ with each other (6 correlations).

\noindent$\bullet$ Equivalent widths of \Lya, \CIV, \HeII\ and \CIII\ with each other (6 correlations).

\section{Emission line ratios}
The relative intensities of emission lines provide a powerful tool to examine the physical state of the extended emission line gas in HzRGs. The (rest-frame) UV emission lines from different elements (mainly hydrogen, carbon, helium, and nitrogen) and the three ionization states of the carbon lines (\CII, \CIII, and \CIV) can provide valuable information on the relative abundance of these elements, and can distinguish between ionization mechanisms. 

An efficient way to compare the data with model predictions is to use line-ratio diagrams. For HzRGs, the rest-frame optical lines which are normally used for such studies have shifted to the near-IR, and we have to use the rest-frame UV-lines that have shifted into the optical. VM97\nocite{vil97} and ADT98\nocite{all98} have used line ratios involving the \CIV, \HeII, \CIII\, and \CII\ lines to compare HzRG spectra (from R\"ottgering 1997\nocite{rot97}) with the predictions from photo-ionization and shock models. 
Our HzRG sample contains almost three times as many sources as the samples in VM97 and ADT98: 58 sources have simultaneously observed \CIV, \HeII, and \CIII\ lines, with redshifts $1.2<z<3.8$, radio sizes $D<365$~kpc, and radio powers $34.5<\log(P_{325})<36.5$. In this section, we shall examine the dependence of the line ratios on these parameters, and on the physical parameters determined from the shock and photo-ionization models. Because we have seen in \S4.4 that all of these parameters have an artificial redshift dependence, we shall first disentangle the mutual dependences of the line ratios on these parameters. After introducing the theoretical models, we shall compare the data with the range of model predictions, and examine the ratios involving \Lya\ and two rest-frame optical lines determined from near-IR spectroscopy.

\subsection{HzRG UV line ratios}
In \S 4.4, we found an increasing dependence of the \CIV, \HeII, \CIII\, and \CII\ line luminosities with redshift. This selection effect will also affect the line ratios. To properly examine the trends involving line ratios, we used a four-parameter Spearman partial rank analysis (Table \ref{UVratiocorrelations}).
\begin{table*}
\begin{center}
\begin{tabular}{lcrrrcrrrcrrr}
Line ratio & & \multicolumn{3}{c}{$z$} && \multicolumn{3}{c}{$D$} && \multicolumn{3}{c}{$P_{325}$} \\
\cline{3-5}
\cline{7-9}
\cline{11-13}
 & Number & $r_{rz}$ & $r_{rz,DP}$ & $\sigma$ && $r_{rD}$ & $r_{rD,zP}$ & $\sigma$ && $r_{rP}$ & $r_{rP,zD}$ & $\sigma$ \\
\hline
  \Lya\ / \CIV & 41 & $-0.13$ & $-0.06$ & $-0.34$ && $ 0.11$ & $ 0.08$ & $ 0.47$ && $-0.09$ & $-0.05$ & $-0.33$ \\
 \Lya\ / \HeII & 32 & $-0.24$ & $-0.06$ & $-0.30$ && $ 0.15$ & $ 0.16$ & $ 0.83$ && $-0.35$ & $-0.32$ & $-1.71$ \\
 \Lya\ / \CIII & 27 & $-0.19$ & $-0.20$ & $-0.93$ && $-0.16$ & $-0.23$ & $-1.10$ && $-0.19$ & $-0.17$ & $-0.81$ \\
 \CIV\ / \HeII & 48 & $-0.10$ & $-0.13$ & $-0.87$ && $-0.05$ & $-0.08$ & $-0.53$ && $ 0.02$ & $ 0.08$ & $ 0.51$ \\
 \CIV\ / \CIII & 43 & $ 0.08$ & $ 0.09$ & $ 0.55$ && $-0.01$ & $-0.00$ & $-0.01$ && $-0.05$ & $-0.06$ & $-0.40$ \\
  \CIV\ / \CII & 11 & $ 0.19$ & $ 0.15$ & $ 0.38$ && $ 0.28$ & $ 0.53$ & $ 1.45$ && $-0.74$ & $-0.80$ & $-2.71$ \\
\CIII\ / \HeII & 45 & $-0.24$ & $-0.25$ & $-1.65$ && $ 0.03$ & $-0.03$ & $-0.18$ && $ 0.03$ & $ 0.09$ & $ 0.57$ \\
 \CII\ / \HeII & 14 & $ 0.32$ & $ 0.19$ & $ 0.57$ && $-0.63$ & $-0.65$ & $-2.31$ && $ 0.31$ & $ 0.58$ & $ 1.97$ \\
 \CII\ / \CIII & 29 & $-0.41$ & $ 0.08$ & $ 0.37$ && $ 0.69$ & $ 0.60$ & $ 3.39$ && $ 0.03$ & $-0.08$ & $-0.39$ \\
\hline
\end{tabular}
\end{center}
\caption[]{Spearman partial rank correlation analysis for the correlations between nine UV line ratios and redshift z, radio size $D$, and radio power $P_{325}$. $r_{rz}$ is the Spearman rank correlation coefficient between the line ratio $r$ of the lines in the first column and $z$, while $r_{rz,DP}$ is the Spearman partial rank coefficient of that correlation, taking the possible correlations with $D$ and $P_{325}$ into account. $\sigma$ is the significance of the partial rank correlation, as in Table 2. The last six columns similarly examine the correlations with radio size and radio power.}
\label{UVratiocorrelations}
\end{table*}

\begin{figure}
\psfig{file=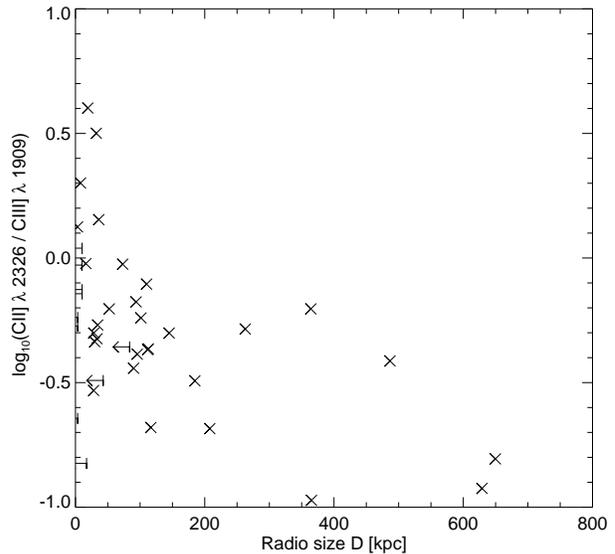,width=8.5cm}
\caption[]{The \CII/\CIII\ ratio plotted versus radio size. Note the absence of large sources with strong \CII.}
\label{CIIoCIIID}
\end{figure}

We find that only one of the correlations with more than 20 elements has more than $2\sigma$ significance, indicating that redshift, radio size or power do not have a strong influence on most emission line ratios.
The only one of these correlations that is significant is the one between the \CII/\CIII\ ratio and radio size $D$, which BRL00\nocite{bes00b} reported as evidence that small sources in their sample of 14 3CR galaxies at $z\sim$1 are ionized by shocks and larger sources by photo-ionization. Our sample, which includes all their sources, is twice as large, but only adds relatively small sources with $D<150$~kpc. The Spearman partial rank analysis shows that the highest probability of correlation of \CII/\CIII\ is indeed with $D$, and not with $z$ or $P_{325}$. The results of BRL00\nocite{bes00b} were not strongly affected by selection effects, because their 3CR subsample has a limited redshift range $0.7<z<1.2$. The addition of six upper limits to the radio size (Fig. \ref{CIIoCIIID}) is consistent with the \CII/\CIII\ - $D$ correlation because no sources $>50$~kpc with \CII\ stronger than \CIII\ are found.
%

Stern \etal (1999) \nocite{ste99} (hereafter S99) reported a possible correlation between the \CIV/\CIII\ ratio and radio power. We find no evidence for a correlation of \CIV/\CIII\ with $P_{325}$, eventhough S99 used 25 of our 43 data points. This is partly due to the exclusion of five sources (of which three are from S99) with only upper limits to the radio size. These three excluded sources are the only ones where \CIII\ is stronger than \CIV. If we retain these five sources and replace their upper limits to the radio size by detections\footnote{As discussed in \S4.2, we cannot use survival analysis on dependent variables.}, we find a Spearman partial rank coefficient $r_{rP,zD}=0.22$ and significance level $D_{rP,zD}=1.35$. Further differences are caused by our different determination of the radio power (\S 3.1.1). It is clear that a larger and better defined sample with multi-frequency radio data to determine the radio power and deep spectroscopy to break the artificial redshift dependence of the line luminosities is needed to examine whether the \CIV / \CIII\ line ratio is correlated with redshift, radio size, or radio power.

To summarize, the only strong correlation between the UV line ratios and radio properties is a dependence of the \CII/\CIII\ ratio on radio size $D$. We find no evidence that the radio power and ionization state are correlated.

\subsection{Shock and photo-ionization models}
Of great importance in the study of the nature of the extended emission line regions in HzRGs is the mechanism responsible for producing the observed ionization state. The most likely mechanisms include photo-ionization from a central AGN, and shock ionization from the radio jets propagating through the interstellar medium. In this section, we describe three theoretical models that can predict the line ratios for these processes.

To calculate the predictions from shock ionization, we use the models of Dopita \& Sutherland (1996; hereafter DS96)\nocite{dop96}. Their models consist of two components: (i) a down-stream shock component, dominated by the radiative cooling of the gas behind the shock, and (ii) an up-stream precursor component, dominated by the photo-ionization by the radiation from the shocked gas. Their models assume solar metallicities, and present a grid of model predictions for shock velocities in the range $150<v<500$~km~s$^{-1}$, and for values of the magnetic parameter $0<B/\sqrt{n}<4 \mu$G~cm$^{3/2}$, with $B$ the pre-shock transverse magnetic field and $n$ the pre-shock number density. This magnetic parameter controls the effective ionization parameter in the down-stream emitting component of the shock since at high shock velocities, the transverse magnetic field limits the compression caused by the shock through a balance of the magnetic pressure of the cloud and the ram pressure of the shock. 
Future versions of these models extend the shock velocities out to 1000~km~s$^{-1}$, and will include non-solar metallicities (\cite{bic00}; Sutherland, private communication).
To derive the \HeIIfull\ fluxes (not included in the tables of DS96) from the \hbox{He~$\rm II$}~$\lambda$~4686 fluxes that are provided, we assume a ratio of 9 (\eg \cite{mac85}). Note that the shock+precursor models of ADT98 do not include this factor for the precursor component so that their \HeIIfull\ values are underestimated by a factors of $\sim 1-6$ (Allen, private communication), whilst VM97 plotted the shock and precursor models separately, which is inappropriate for HzRG spectra, since these are spatially integrated over both the shock and precursor gas.

For the photo-ionization models, VM97 use the MAPPINGS~I code developed by Binette, while ADT98 use the MAPPINGS~II code developed by Sutherland. The differences between the two model predictions are minor for the relevant range of parameters. We use CLOUDY version C94 (\cite{fer96}), and also find very similar results, providing another independent consistency check with the results of VM97 and ADT98.
We calculate the model spectra for the same range of parameters as ADT98 to facilitate comparison between the predictions of the different codes. We assume solar metallicities, and an ionizing continuum which is a power law spectrum ($\phi_{\nu} \propto \nu^{\alpha}$) with lower and higher energy cutoffs of 0.01~Ryd and 100~Ryd (1.36~keV), but the models do not depend strongly on the exact values of these limits. We calculate spectra for a power law spectral index $\alpha=-1.5$ which well matches the low redshift Seyfert spectra (\eg \cite{eva99}), and for $\alpha=-1.0$, which VM97 find to match the HzRG data better. For the hydrogen density, we used a value of $n=100$, as commonly used for the extended emission line regions (\eg\ \cite{mcc90b}) and a high density value of $n=1000$~cm$^{-3}$. We vary the ionization parameter\footnote{Defined as $U=(c n)^{-1} \int_{\nu_0}^{\infty}(\phi_{\nu}d\nu)/h\nu$, with $\phi_{\nu}$ the monochromatic ionizing energy flux impinging on the cloud, $\nu_0$ the ionizing potential of hydrogen, $n$ the total gas density at the front face of the cloud, $c$ the speed of light, and $h$ Planck's constant.} $U$ from 0.001 to 0.1 in steps of 0.5 dex to cover the entire range of observed line ratios.

Binette, Wilson \& Storchi-Bergmann (1996, hereafter B96)\nocite{bin96} present an alternative photo-ionization sequence by considering emission from two distinct cloud populations. In their model, the light from the photo-ionizing source first passes through a population of matter-bounded (optically thin) clouds, and subsequently strikes ionization-bounded (optically thick) clouds located further outwards. They produce a sequence by varying the the parameter \AMI, defined as the ratio of the solid angle subtended by the matter-bounded clouds to the solid angle subtended by the ionization-bounded clouds. They keep the power law spectral index of the ionizing continuum incident at the matter-bounded clouds constant at $\alpha=-1.3$ and the ionization parameter constant at $U_{MB}=0.04$. By adding this ionization-bounded component, B96 solve several shortcomings of the classical photo-ionization models, like the inability to produce strong high excitation UV lines for reasonable values of $U$ inferred from the optical lines. Because the \AMI\ models require that the ionizing continuum incident on the ionization-bounded clouds first has to be ``filtered'' by the matter-bounded clouds, the values of \AMI\ cannot strictly be less than unity. However, if the matter-bounded clouds are obscured along the line of sight, we can observe apparent \AMI$<$1 values.

\subsection{Comparison with the observations}
Figure \ref{CandHediagnostics} reproduces three diagnostic diagrams from VM97 and ADT98, showing the \CIV, \HeII, and \CIII\ lines, and a Carbon-only plot from ADT98, showing the ratios of \CIV, \CIII, and \CII.
\begin{figure*}
\psfig{file=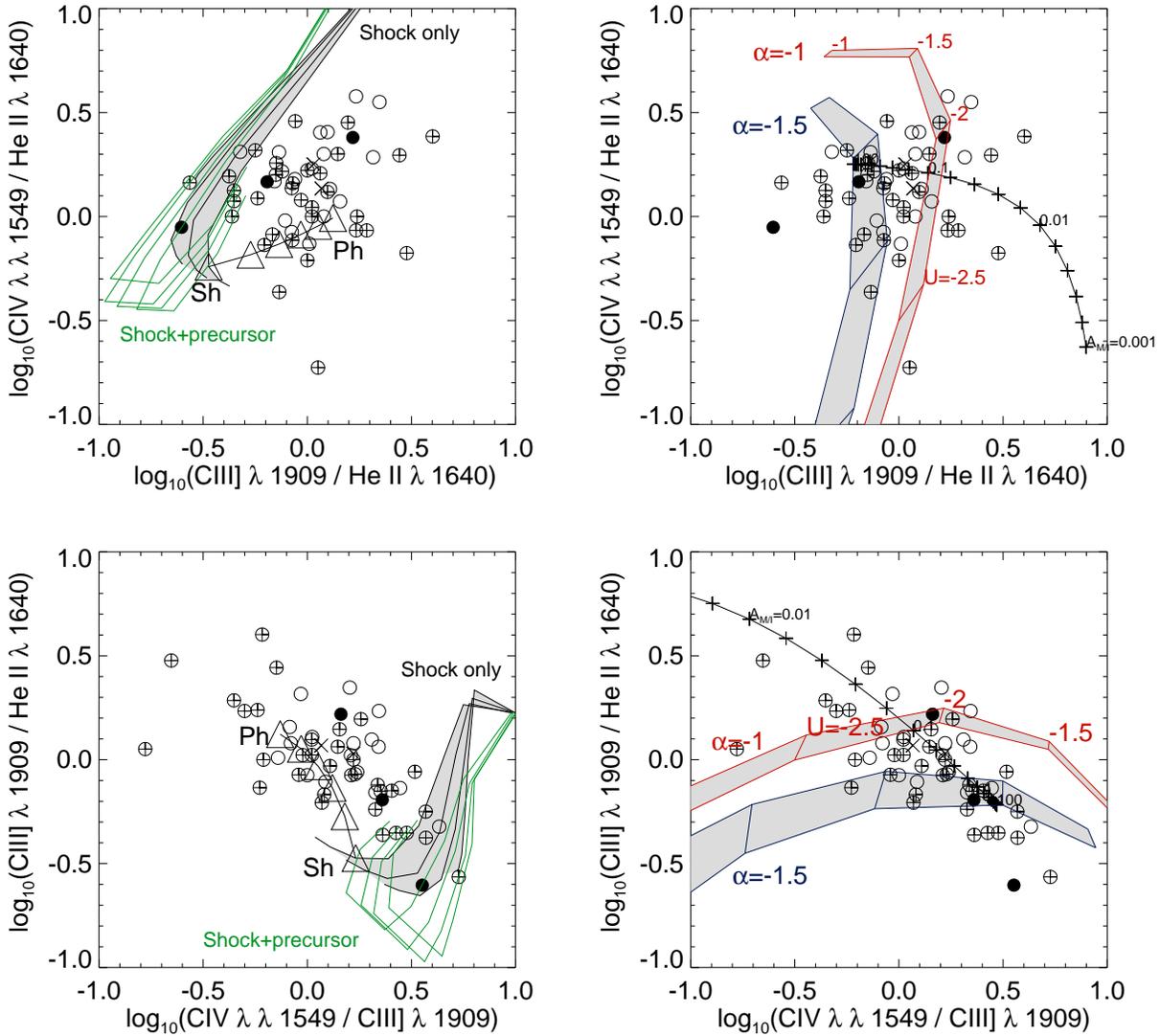,width=16.4cm}
\caption[]{Line ratio diagnostic diagrams for \CIV, \HeII, \CIII, and \CII. The left panels show the shock (shaded grid) and shock+precursor (unshaded grid) models. Shock velocity increases along the lines from 150~km~s$^{-1}$ at the top or right to 500~km~s$^{-1}$ at the end of the curves, which show four values of the magnetic parameter $0<B/\sqrt{n}<4~\mu$G~cm$^{3/2}$. 
The sequence shown by open triangles represents a composite model of a pure shock model with $B/\sqrt{n}=4~\mu$G~cm$^{3/2}$, $v=$400~km~s$^{-1}$ and a photo-ionization model with power-law spectral index $\alpha=-1.0$, hydrogen density $n=100$~cm$^{-2}$ and ionization parameter Log$_{10}(U)=-2.25$. The triangles represent different fractions of each model in steps of 20\% (see text for details).

The right panels show the photo-ionization models. We show four photo-ionization sequences, with the values of log$_{10}(U)$ every 0.5 dex. The top shaded grid represents two models with power law spectral index $\alpha=-1$, the bottom grid models are for $\alpha=-1.5$. The boundaries of the shaded sequences are models for a hydrogen density $n=100$ (left or bottom) and $n=1000$ (top or right). The single curve is the \AMI\ sequence of B96, with values $0.01<$\AMI$<100$ and tickmarks every 0.2 dex. 

The data are repeated in both panels. Open circles represent sources at $z<2$, circled +-signs at $2<z<3$, and filled circles at $z>3$. The asterisks represent the two nearby Seyfert galaxies observed by Evans \etal (1999)\nocite{eva99}.}
\label{CandHediagnostics}
\end{figure*}
\begin{figure*}
\psfig{file=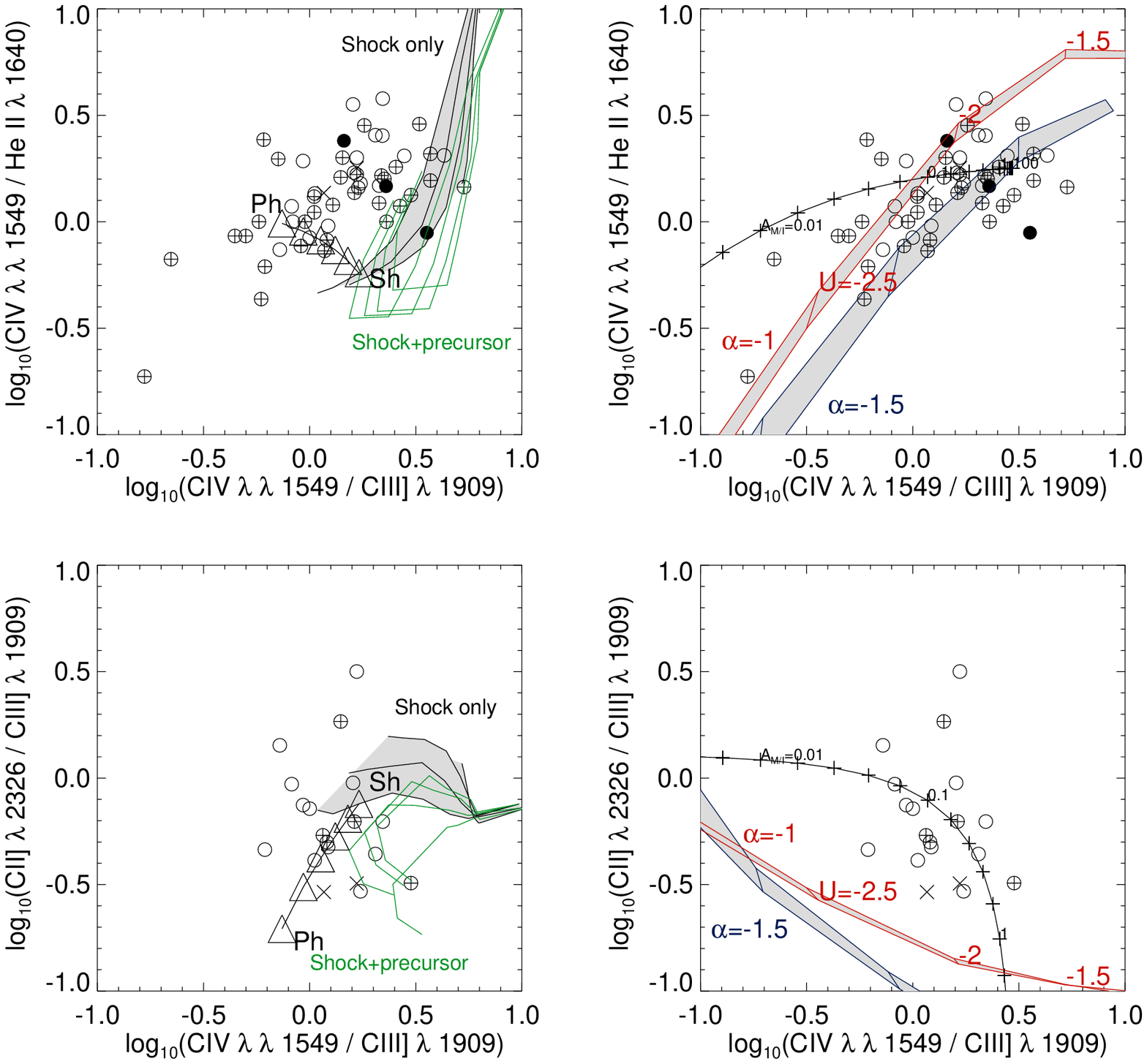,width=16.4cm}
\addtocounter{figure}{-1}
\caption[]{, continued.}
\end{figure*}
%

\subsubsection{Shock models}
From the left panels in Figure \ref{CandHediagnostics}, we find that the present shock or shock+precursor models cannot explain the ratios involving \CIV, \HeII, and \CIII, while roughly half of the HzRGs in the Carbon-only plot coincide with the highest shock velocity (500~km~s$^{-1}$) models. In the plots without \CII, the least discrepant predictions are the highest shock velocity shock+precursor models. The disturbed kinematics of the emission line gas indicate that shocks with velocities $>$500~km~s$^{-1}$ are probably occurring in HzRGs (\cite{bic00}), so it will be of great interest to compare the data with higher shock velocity models (Sutherland \etal, in preparation). Preliminary results indicate that the existing models cannot be simply extrapolated (Allen, private communication). Moreover, the DS96 shock models have only been calculated for solar metallicities, which are unlikely to be correct in the extended emission line regions at the highest redshifts (\S 6.3, \cite{bin00}). Lower metallicities will lead to less efficient cooling and raise the temperature in the precursor area, leading to stronger emission. We conclude that the present shock models can only reproduce the \CII\ vs. \CIII\ ratio, but that extensions of the present models to higher shock velocities and lower metallicities will be needed to fully determine the importance of this ionization mechanism.

\subsubsection{Pure photo-ionization}
From the right panels in Figure \ref{CandHediagnostics}, we find that pure photo-ionization provides good fits to the \CIV/\HeII\ and \CIV/\CIII\ ratios, and a reasonable fit to the \CIII/\HeII\ ratio, but fails to explain the \CII/\CIII\ ratio. In the plots without \CII\, the majority of the data-points are bracketed by the models with power law spectral indices of the incident ionizing continuum of $\alpha=-1$ and $\alpha=-1.5$. In the plot involving \CII\, the $\alpha=-1$ model provides the least discrepant predictions. All plots suggest ionization parameters $-2.5 \simlt$ log$_{10}$(U) $\simlt -2$ for the $\alpha=-1$ models and $-2 \simlt$ log$_{10}$(U) $\simlt -1.5$ for the $\alpha=-1.5$ models. 

The poor fit in the Carbon-only plot indicates that the different ionization stages of Carbon originate from distinct regions in the galaxies. The fact that the \AMI\ sequence provides the best fit in this diagram (see below) is consistent with this idea, as in this model we observe simultaneously emission from clouds with different incident ionizing radiation. We return to this in \S6.1.

\subsubsection{\AMI\ sequence}
The predictions for the \AMI\ sequence in the models of B96 provide good fits to the data, but fails to reproduce the sometimes large spread of the points. This could be solved by changing the dust content, metallicity, spectral index or ionization parameter $U_{MB}$ of the ionizing continuum incident on the matter-bounded clouds, or by internal reddening\footnote{Galactic reddening will be negligible because most HzRGs have been identified from samples avoiding the Galactic plane, and some of the spectra in the literature have been corrected for reddening.}. 
A value of \AMI$\sim$0.1 seems to best fit the data in all four diagrams, although the scatter ranges from \AMI$\sim$0.02 to \AMI$\simgt$100. A value of \AMI$\sim$0.1 means that as much as 90\% of the matter-bounded clouds has to be obscured along the line of sight. In low redshift Seyferts, B96 and Evans \etal (1999)\nocite{eva99} found values of \AMI\ slightly higher than unity provided the best fit to the data. If the \AMI\ sequence is the correct ionization mechanism model, the matter-bounded clouds would be much more obscured at higher redshift, or more obscured in radio galaxies than in Seyferts.
\begin{figure*}
\psfig{file=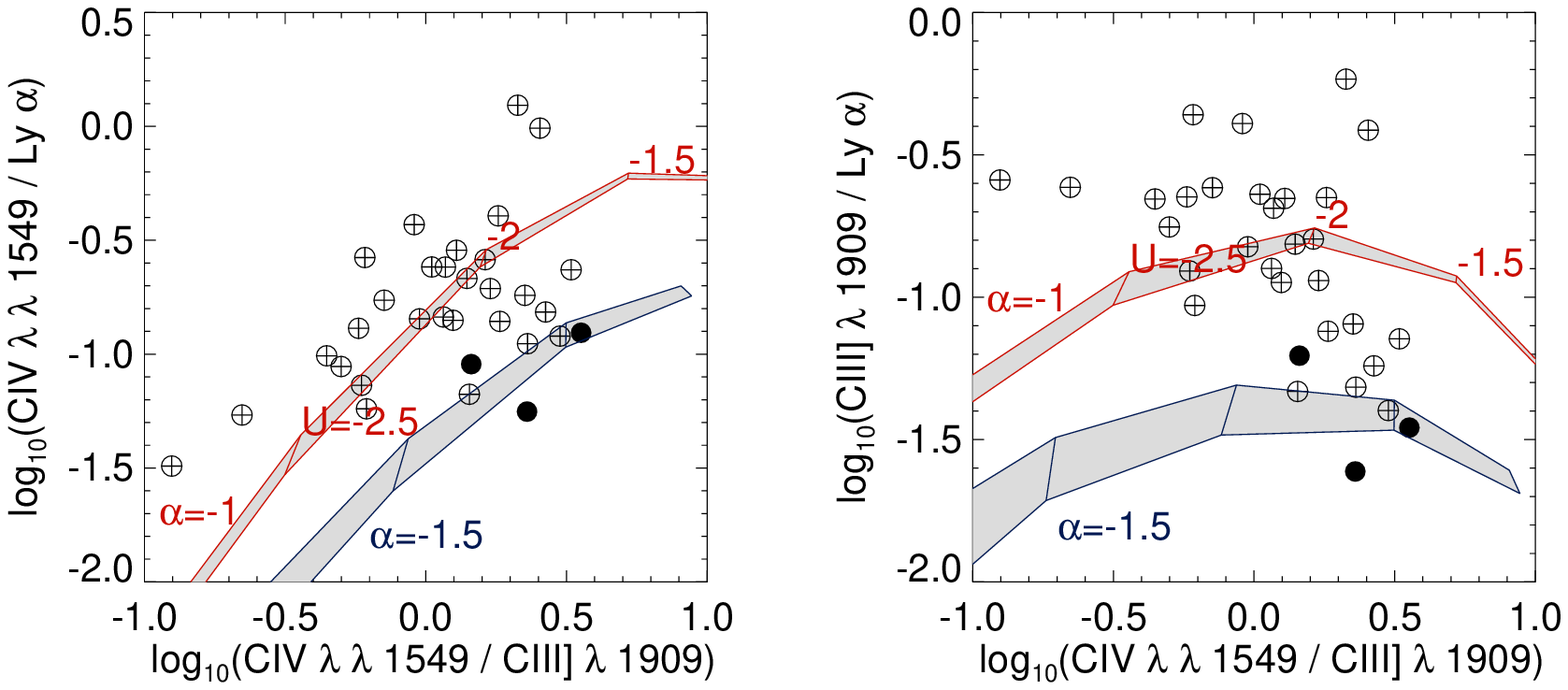,width=16.5cm}
\caption[]{Line ratio diagnostic diagrams for \Lya, \CIV\ and \CIII. Models are as in Figure \ref{CandHediagnostics}. Note the systematical over-luminosity of \Lya\ at $z>3$ (filled circles).}
\label{Lyadiagnostics}
\end{figure*}
%

A possible way to obscure these regions could be preferential obscuration by dust. Dust masses up to $10^9$M$_{\odot}$ have been detected in HzRGs (\eg \cite{arc00}). If these large amounts of dust are located near the central parts of HzRGs, they could lead to obscuration of either the matter-bounded or ionization-bounded clouds. There are six radio galaxies in our sample with solid submm detections which also have published \CIV, \HeII\ and \CIII\ measurements (4C41.17, Chini \& Kr\"ugel 1994\nocite{chi94}, Dunlop \etal 1994\nocite{dun94}; MG~J1019+0534, Cimatti \etal 1998a\nocite{cim98a}; 4C24.28, 4C28.58 and 4C48.48, Archibald \etal 2000\nocite{arc00}; TN~J0121+1320, Reuland \etal, in preparation). These six sources are not preferentially located in any part of the diagrams, and do not fall in the low \AMI\ region. We conclude that dust obscuration of the matter-bounded clouds is an unlikely explanation of the very low \AMI\ values. A variation of the \AMI\ sequence by changing the ionization parameter or power law spectral index incident on the matter-bounded clouds would be required to explain the high obscuration of the matter-bounded clouds and the observed scatter with these models.

\subsubsection{Summary of comparison with the data}
To summarize, we find that using only the line-ratio diagrams involving \CIV, \HeII, and \CIII, the pure photo-ionization models provide the best fit to the HzRG data, but these models clearly fail to reproduce the observed \CII / \CIII\ ratio, which can be well fit by high shock velocity models. The models that use a combination of matter-bounded and ionization-bounded clouds seem to provide a reasonable fit to all four lines, but the spread of the points does not seem to follow the \AMI\ sequence, and suggests other parameters dominate the intrinsic differences in the ionization levels of the individual HzRGs. However, the combination of at least two zones with different ionization continua seems to be required to explain the emission line ratios in \CIV, \HeII, \CIII, and \CII. 

In the remainder of this section, we shall compare these model predictions with some less commonly used emission line ratios. In \S6.1, we shall return to the multiple zone ionization mechanisms. 

\subsection{Diagnostic diagrams with \Lya}
In almost all HzRGs, the brightest UV line is \Lya. This line has not been frequently used in diagnostic line-ratio diagrams, because it is highly sensitive to resonance scattering inside and around the excited clouds, and to absorption by dust.

Villar-Mart\'\i n, Binette \& Fosbury (1996)\nocite{vil96} have examined the effects of resonance scattering and dust on UV lines in HzRG. They found that cases where the \Lya\ emission is weak relative to the other UV lines can better be explained by geometrical (viewing angle) effects rather than by large amounts of dust, which would also affect the other UV lines, notably \CIV. 
Another factor affecting the total flux in \Lya\ is absorption by associated \HI. In \S 3.2.2, we saw that this \HI\ absorption will reduce the total flux of the \Lya\ line by no more than 60\%, which in most cases will not completely destroy the diagnostic value of the \Lya\ flux for line ratios. In this section, we shall therefore first compare the observed ratios involving \Lya\ with the values of the photo-ionization models found from the other UV lines, and then consider the geometrical and dust effects.
\begin{figure*}
\psfig{file=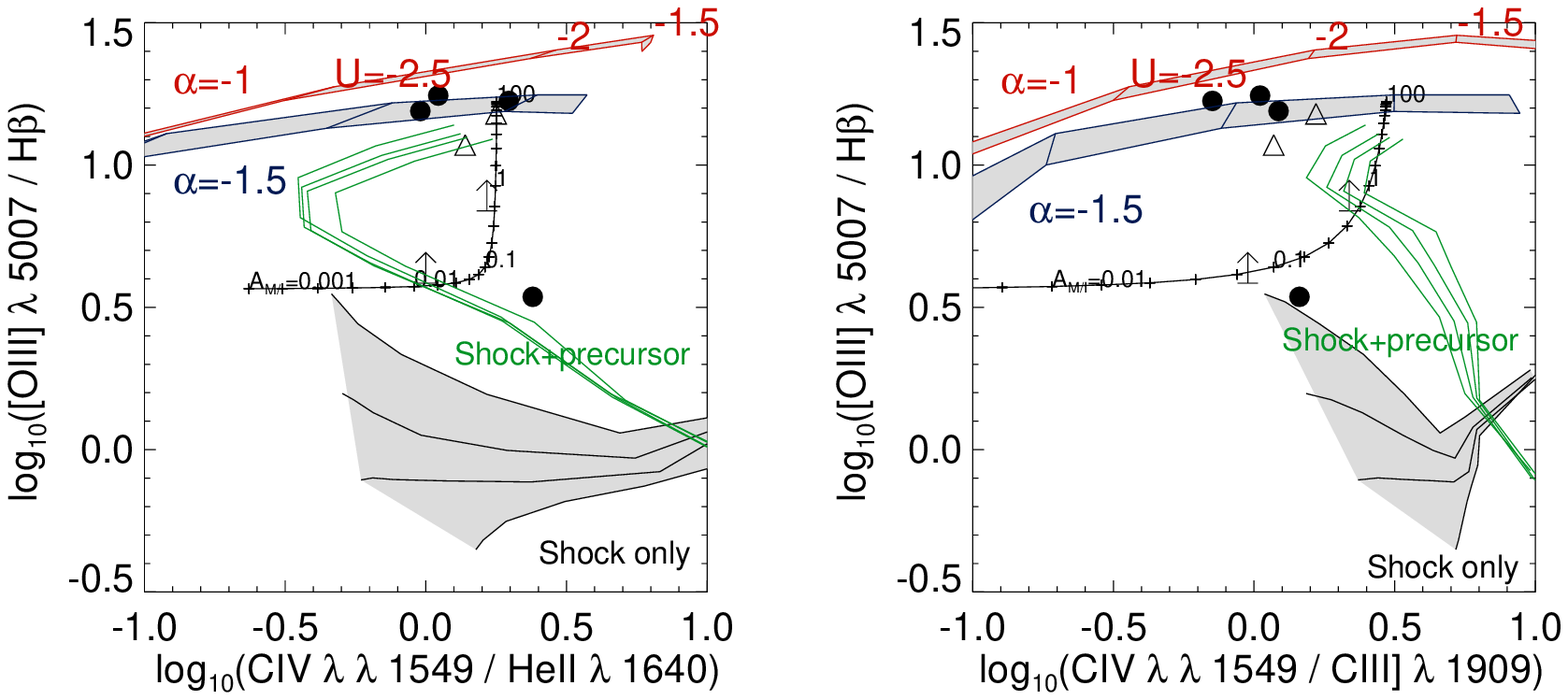,width=16.5cm}
\caption[]{Optical-UV line ratio diagnostic diagrams. Dots and lower limits are HzRG data, and open triangles are the Seyfert galaxies NGC ~5643 and NGC~5728 (\cite{eva99}). Models are as in Figure \ref{CandHediagnostics}.}
\label{UVoptdiagnostics}
\end{figure*}
%

Figure \ref{Lyadiagnostics} presents two diagnostic diagrams with line-ratios of \Lya, \CIV, and \CIII. We find that the simple photo-ionization model with a power law spectral index $\alpha=-1$ provides a good fit to the data in both diagrams. The range of the ionization parameter $-2.5 <$log$_{10}$(U)$<-2$ is consistent with the values found from the diagrams in Fig. \ref{CandHediagnostics}. Contrary to the diagrams in Fig. \ref{CandHediagnostics}, the $\alpha=-1.5$ models do not provide a good fit to the data, because they overestimate the \Lya\ flux. This might be due to geometrical effects: changing the viewing angle from front to back illuminated \Lya\ can increase the \CIV/\Lya\ ratio by almost an order of magnitude (see Fig. 7 of Villar-Mart\'\i n \etal (1996)\nocite{vil96}. We therefore do not take this better fit as evidence for a flatter spectral index of the incident ionizing continuum.

Some individual HzRGs have highly discrepant \Lya/\CIV\ and \Lya/\CIII\ ratios. Both under- and over-luminous \Lya\ occurs. Dust extinction has been proposed as an explanation for under-luminous \Lya\ in the two most discrepant objects in our HzRG sample, TX~0211$-$122 at $z=2.34$ (\cite{oji94}) and MG~J1019+0534 at $z=2.765$ (\cite{dey95}, \cite{cim98a}). The \Lya\ in these objects is a factor of $\sim$3 lower compared to the best fitting photo-ionization models and the bulk of the other HzRGs. However, in two objects with equally large amounts of dust, 4C~41.17 at $z=3.8$ and TN~J0121+1320 at $z=3.516$, the \Lya\ is a factor $\sim$3 brighter than the models and the mean of the other HzRGs. TN~J0205$+$2242 at $z=3.506$ does not have a large dust mass detected, while it is also $\sim 4 \times$ over-luminous in \Lya. We conclude that there is no strong correlation between the detection of a large global dust mass and an anomalously low \Lya\ flux. If dust obscuration is an important process in suppressing the \Lya\ luminosity, it would have to be a localized process, but this would be hard to achieve, given the large extent of the \Lya\ emission. High resolution \Lya\ imaging, combined with sensitive imaging of the IR-dust emission (with ALMA) would be needed to examine this in more detail. 
Metallicity differences and geometrical effects probably play a more important role than dust obscuration in the destruction of \Lya. The most remarkable observation from Figure \ref{Lyadiagnostics} is that three of the seven HzRGs with over-luminous \Lya\ in the \CIII/\Lya\ versus \CIV/\CIII\ plot are at $z>3$. We shall return to this in \S 6.3.

\subsection{UV-optical diagnostic diagrams}
In HzRGs, the rest-frame optical emission lines which are commonly used in line ratio diagnostic diagrams at low redshift, are redshifted into the near-IR. It has been a big technological challenge to obtain near-IR spectra of HzRGs with 2-4m telescopes, but recently, near-IR spectroscopy of HzRGs has become feasible with the advent of efficient near-IR spectrographs on 8-10m class telescopes (\eg \cite{lar00}). It is to be expected that more near-IR spectroscopy of HzRGs will be available soon. To date, a total of 15 $z>2$ radio galaxies have published near-IR spectra obtained with the 3.8m UKIRT and 2.2m UH telescopes (\cite{eal93}, \cite{iwa96}, \cite{eva98}), and recently the first near-IR spectrum obtained with NIRSPEC at Keck was published (\cite{lar00}). 
As shown by ADT98, a diagnostic diagram of the optical \OIII\ to \Hbeta\ versus the UV \CIV/\HeII\ or \CIV/\CIII\ line ratios can separate the shock and shock+precursor models. Such diagrams have the added advantage that they are relatively insensitive to the effects of dust extinction, as the ratios are determined from lines close in wavelength. In four objects (3C 256, MRC~2025-218, 4C~40.36 and 4C~41.17), \Hbeta\ was detected as well as \OIII, and two more objects (USS~0828+193 and 4C~48.48) have lower limits to the \OIII\ to \Hbeta\ ratio. The uncertainties in these points will be higher than in the optical spectroscopy because of the low signal-to-noise ratio and resolution of the near-IR spectroscopy.

Figure \ref{UVoptdiagnostics} reproduces the line-ratio diagrams from ADT98 with the five HzRG data points added. We also compare the HzRG data with combined HST and ground-based spectroscopy of two low-redshift Seyferts (\cite{eva99}). The most firm conclusion from Figure \ref{UVoptdiagnostics} is the exclusion of the pure shock models. Only 4C~41.17 falls in a region of the plot that cannot be explained by pure photo-ionization models, but the \Hbeta\ detection is marginal (\cite{eal93}). The other detection and the lower limits are consistent with the Seyfert points, and fall in a region where the simple photo-ionization models overlap with the highest values of the \AMI\ sequence and the high shock velocities of the shock+precursor models. A comparison with the range of parameter values of the UV diagnostic diagrams suggests that the simple photo-ionization models have more consistent values than the \AMI\ sequence, because the weak \Hbeta, if confirmed by higher S/N observations, requires a higher fraction of matter-bounded clouds than the UV diagrams suggest. 
Alternatively, the highest shock velocity (500~km~s$^{-1}$) predictions of the shock+precursor models can also explain most of the points in Figure \ref{UVoptdiagnostics}. This again would require an extension of these models to higher shock velocities to fully examine the importance of this ionization mechanism in HzRGs.

To summarize, the comparison of the UV and optical line-ratios excludes the pure shock models, and is consistent with the high shock velocity shock+precursor models and the high values of the \AMI\ sequence. Pure photo-ionization models with a power-law spectral index $\alpha = -1$ and ionization parameter $U \sim 0.01$ seem to provide the most consistent fit to HzRG spectra. Future near-IR spectroscopy of HzRGs will be required to differentiate more unambiguously between these models.

\subsection{Nitrogen overabundance}
In the majority of HzRGs, the \NV\ line was not detected. There are, however, several exceptions where \NV\ is strong, and even a few cases where it is as bright as \Lya\ (\eg \cite{oji94}). Neither shock ionization nor variations of the densities or ionization parameters in the photo-ionization models are able to produce such strong \NV\ emission (\cite{vil99c}). 
The most likely explanation is that \NV\ is over-abundant compared to the other species in HzRGs (\cite{oji94}). Fosbury \etal (1998, 1999)\nocite{fos98}\nocite{fos99} found that HzRGs follow a close correlation in a \NV/\HeII\ vs. \NV/\CIV\ diagram which is parallel to the relation defined by the broad line regions of quasars (\cite{ham93}). These authors showed that this sequence can be explained by a large variation of the metallicity (from $Z \sim Z_{\odot}$ to $Z \simgt 10 Z_{\odot}$) caused by rapid chemical evolution in a stellar population composed of massive stars. The parallel sequence defined by the extended emission line regions in HzRGs suggests that we also observe such a metallicity variation on much larger scales than the small central broad line region (\cite{fos99}). Vernet \etal (1999)\nocite{ver99} found that the radio galaxy sequence is best modeled with a metallicity sequence in which the nitrogen abundance increases quadratically, while the other elements increase linearly.

Figure \ref{NVoCIV} presents the \NV/\HeII\ vs. \NV/\CIV\ diagram with all the points from our HzRG sample, of which two thirds have only upper limits for \NV. The inclusion of such a large number of upper limits could lead to redshift selection effects. We believe this is a major problem in our sample, as \NV\ is only observable over a limited redshift interval ($z \simgt 2$), and the highest redshift objects have been observed longer or with larger aperture telescopes. This is also obvious in the flux-flux diagram of the \NV\ and \CIV\ lines, which shows a constant spread of the detections and upper limits on \CIV\ flux.

\begin{figure}
\psfig{file=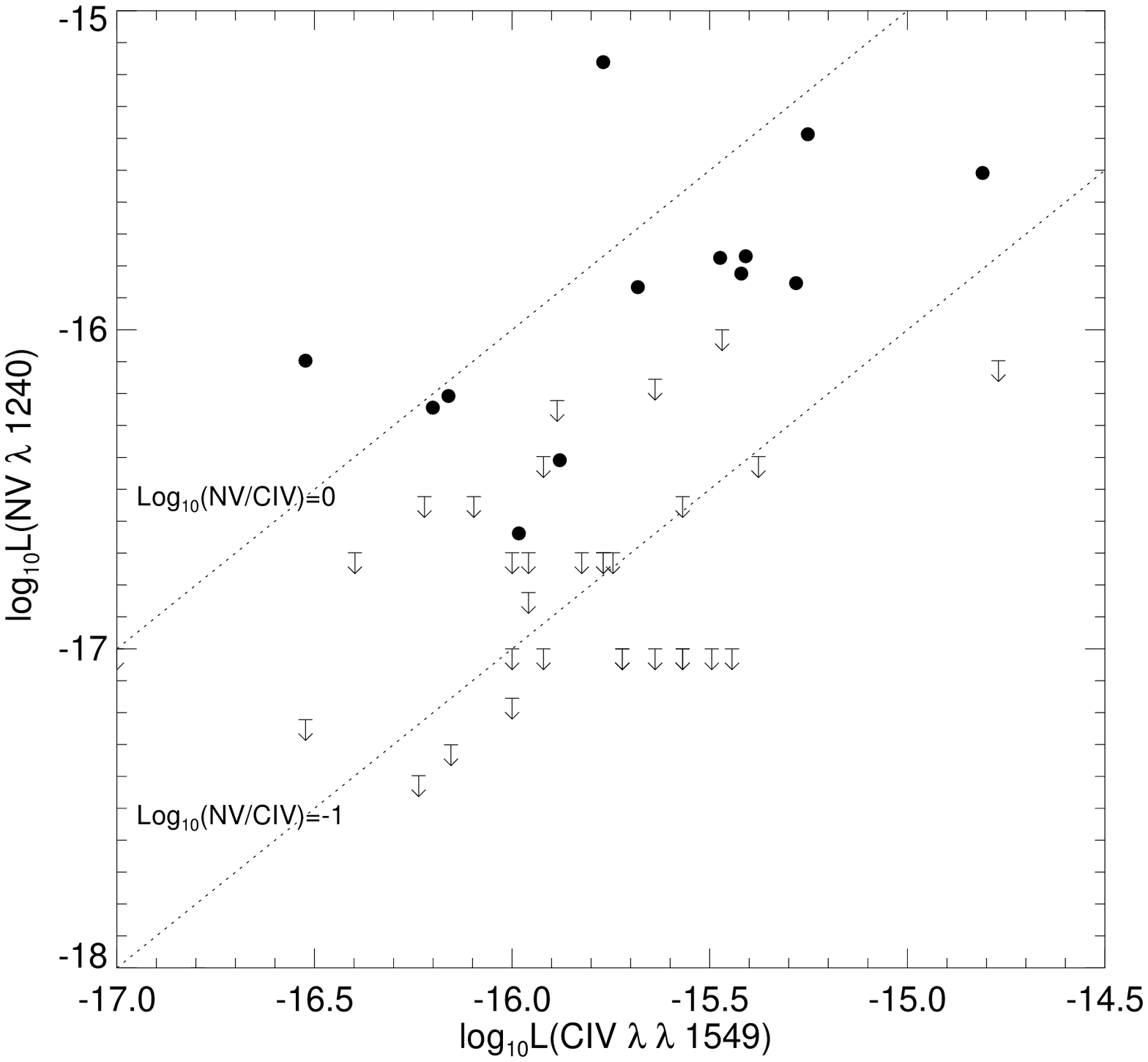,width=8.5cm}
\caption[]{\NV/ flux plotted \CIV\ flux. The diagonal lines indicate constant \NV/\CIV\ ratios. Note that there is no strong clustering of the upper limits at high or low \CIV\ fluxes, which would lead to an artificial correlation in the \NV/\CIV\ ratios.}
\label{NVvsCIV}
\end{figure}
%

\begin{figure}
\psfig{file=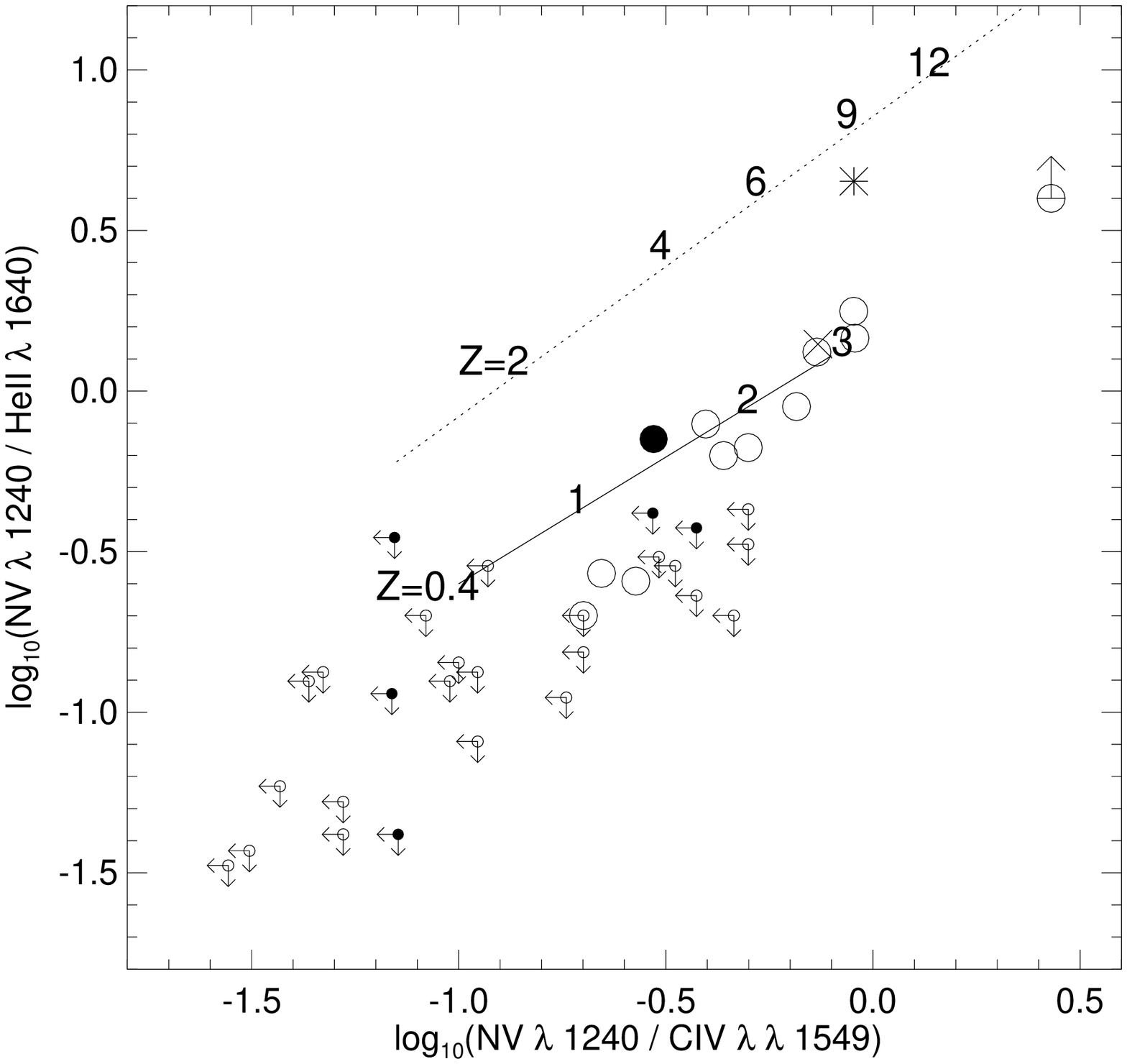,width=8.5cm}
\caption[]{\NV/\HeII\ vs. \NV/\CIV. Radio galaxies at $z<3$ and $z>3$ are represented with open and filled symbols, respectively. The dotted line represents the metallicity sequence defined by quasars (\cite{ham93}), with the numbers along the line representing the metallicity in solar units. The solid line represents a metallicity sequence with quadratic N enhancement (U=0.035, power law spectral index $\alpha=-1.0$) from Vernet \etal (1999)\nocite{ver99}. The starburst galaxy IRAS~F10214+4724 (\cite{ser98}) is indicated with a cross and the hyper-luminous, gravitationally lensed object SMM~J02399$-$0136 (\cite{vil99c}) is indicated by a star. Note that the $z>3$ objects seem to fall preferentially in the low metallicity region of the sequence.}
\label{NVoCIV}
\end{figure}
%

Using survival analysis, we find that the probability that the \NV/\CIV\ and \NV/\HeII\ ratios are correlated with redshift is $<$35\%, while the ratios are strongly correlated with each other (the generalized Spearman rank correlation coefficient is $r=0.95$, signficance level $>99.99\%$). The extension of the radio galaxy sequence to both sides in Fig. \ref{NVoCIV} suggests that the nitrogen abundance shows an even larger variation than the $Z=0.4Z_{\odot}$ to $Z=3Z_{\odot}$ sequence proposed by Vernet. The most extreme source in our sample is TXS~J2353$-$0002 (De~Breuck \etal 2000b\nocite{deb00b}), which has \NV\ emission that is more than twice as luminous as \CIV\ and undetected \HeII\ emission, at least four times weaker than \NV.

Eventhough there is no statistical significance for a correlation between the \NV/\CIV\ ratio and redshift within the $2 \simlt z \simlt 4.4$ redshift interval we can observe both these lines, it is clear from Figure \ref{NVoCIV} that none of the $z>3$ radio galaxies occupy the high metallicity end of the diagram. Moreover, the limits from the VLT and Keck spectra on the \NV\ of TN~J1338$-$1942 at $z=4.11$ and 6C~0140+326 at $z=4.41$ are quite strong (\NV/\CIV\ $<0.07$), and suggest the $z>4$ galaxies have even lower metalicities. It appears that the metalicities in the extended emission line regions of HzRGs show large variations, from values well below solar to several times solar. The highest metalicities only occur in $z \simlt 3$ radio galaxies, while the most distant objects generally have sub-solar metalicities. 

\section{Discussion}
\subsection{Simultaneous shock and photo-ionization}
From the diagnostic diagrams of the UV line ratios, and the combined UV and optical line ratios in \S5, we found that the line-ratios involving \CIV, \HeII, and \CIII\ can best be explained by photo-ionization models, whilst the \CII\ to \CIII\ ratio is better fitted by shock models with high shock velocities. Models that invoke a variation in the ratio of matter-bounded and ionization-bounded clouds provide a more consistent fit to all the UV line-ratios, although the large scatter of the data around the model predictions does not appear to follow the \AMI\ sequence.

We find that within the ranges of emission line fluxes measured in our HzRG sample, neither redshift, nor radio luminosity are important causes of this scatter. The radio size D also does not appear to influence the emission lines, with the notable exception of the \CII/\CIII\ ratio. 
This was interpreted by BRL00\nocite{bes00b} in a \CII/\CIII\ versus \NeIII/\NeV\ diagram as evidence that the ionization mechanism in smaller radio sources is shocks, while the larger sources are photo-ionized. However, the only object from their sample that also has a published \CIV\ flux, 3C~324, lies in the shock+precursor region in their diagram, but cannot be explained by any of the shock models in our diagrams involving the \CIV, \HeII, and \CIII\ lines (Fig. \ref{CandHediagnostics}). The fluxes of these three lines in 3C~324 are very similar, putting the object consistently on the photo-ionization sequence with $\alpha=-1.0$, and $\log_{10}U\approx -2.2$ in all three plots involving these lines. 

We interpret this apparent inconsistency as an indication that the integrated spectra of this, and most likely also other HzRGs, are a composite of differently ionized areas. The lower ionization state \CII\ line is much more sensitive to shock ionization than the other UV lines: in the shock ionization models of DS96, the \CII\ flux is similar to the \CIII\ flux, while in the pure photo-ionization models, \CII\ is $\sim 5\times$ weaker than \CIII. An increasingly more dominant shock-ionized area will thus manifest itself first through brighter low-excitation lines such as \CII. We can now also understand why the \AMI\ sequence provides a much better fit to the Carbon-only plot than the pure photo-ionization models: from Table 2 of B96, we find that 25$\times$ more \CII\ is produced in the matter-bounded clouds than in the ionization-bounded clouds; the addition of the ionization-bounded clouds could mimic the addition of a more dominant shock component. 

To examine this idea of a mixture of shock and photo-ionized gas in the integrated spectra in more detail, we have over-plotted a sequence of the fraction of shock to photo-ionization on the line-ratio diagrams in Figure \ref{CandHediagnostics}. We picked the pure photo-ionization model with $n=100$~cm$^{-3}$, $\alpha=-1.0$ and $\log U=-2.25$, and added a contribution of the shock model with $v=400$~km~s$^{-1}$ and $B/\sqrt{n}=4 \mu$G~cm$^{3/2}$ in steps of 20\%. This approach is likely to be a serious oversimplification, because we expect a range of parameters for both photo-ionization and shock models, but it might be a reasonable approach in a single object, such as 3C~324. 
A 40\% shock contribution, in this object would provide an excellent fit in the Carbon-only plot, while it will not shift the predicted ratios in the other diagrams by as much. The \CII/\CIII\ and \CIII/\HeII\ ratios show the largest variation when the fraction of shock ionization is increased, with the \CII\ flux increasing faster towards the shock models than the \HeII\ flux. A varying shock contribution might well explain some of the scatter observed in the diagrams of Figure \ref{CandHediagnostics}. Assuming that the $\alpha=-1.0$ pure photo-ionization sequence is the most representative, the scatter in all diagrams always seems to fall along the side of the sequence pointing towards the (high shock velocity) shock models. Because the \CII\ line is the most sensitive to shock ionization, most points in the Carbon-only plot do not fall on the photo-ionization sequence, but closer to the shock models.

Another effect which could raise the \CII/\CIII\ ratio is differential extinction. For a single, central photo-ionizing source (the AGN), the ionization parameter $U$ of a cloud located closer to the source will be higher than for the same cloud located further out. If the higher excitation photo-ionized gas is more centrally located than the lower excitation gas, it will be more subject to extinction. However, the \CII/\CIII\ ratio is not very sensitive to changes in the ionization parameter $U$ for an ionizing continuum with an $\alpha=-1.0$ power law spectrum. For example, to change the ionization parameter from $\log U=-2.0$ to $\log U=-2.5$ we need to move three times further out from the AGN, while the \CII/\CIII\ ratio only changes by a factor of two. We consider it very unlikely that this effect can be dominant, and so we neglect it.

The above findings could undermine the diagnostic value of the different line-ratio diagrams. If the integrated HzRG spectra are a composite of different regions, we should not necessarily try to fit all lines with a single model. Photo-ionization remains the main process in HzRGs, but using only high excitation lines, we cannot exclude shock ionization, like VM97 and ADT98 did. The high excitation diagrams, which include lines that are insensitive to shock ionization, can be used to determine the best fitting parameters of the photo-ionization model, which can then subsequentially be used in the diagram which include lines that are more sensitive to shock ionization, such as \CII, to estimate the relative contribution of shock ionization. 

The study of BRL00\nocite{bes00b} has shown that significant shock contributions only occur in sources with radio sizes $\simlt 150$~kpc. Because only 15\% of the $z>2$ galaxies in our HzRG sample have radio sizes $>$150~kpc (and only 6\% $>$200~kpc), the contribution of shock ionization in HzRGs might well be important. There is also further kinematical and morphological evidence for the presence of shocks in HzRGs. At relatively low redshifts ($z \simlt 1$), the line widths in radio galaxies are $\simlt 500$~km~s$^{-1}$ and can also be explained by gravitational origins (\cite{bau00}). However, at $z\simgt 2$, the line widths of the UV lines are $\sim 1500$~km~s$^{-1}$, which is more easily explained by acceleration due to the passage of a bow shock associated with the expansion of the radio source, although different processes might also play a role, such as (i) infall of material from large distances, (ii) large scale outflows of companion Lyman break galaxies, or (iii) bipolar outflows produced by super-winds (\cite{vil99a}). In some objects at low redshift (\eg \cite{wvb85}, \cite{vil99b}) and at high redshift (\cite{bic00}), there is also morphological evidence for the interaction of the radio jet with the ambient gas. 
\begin{figure}
\psfig{file=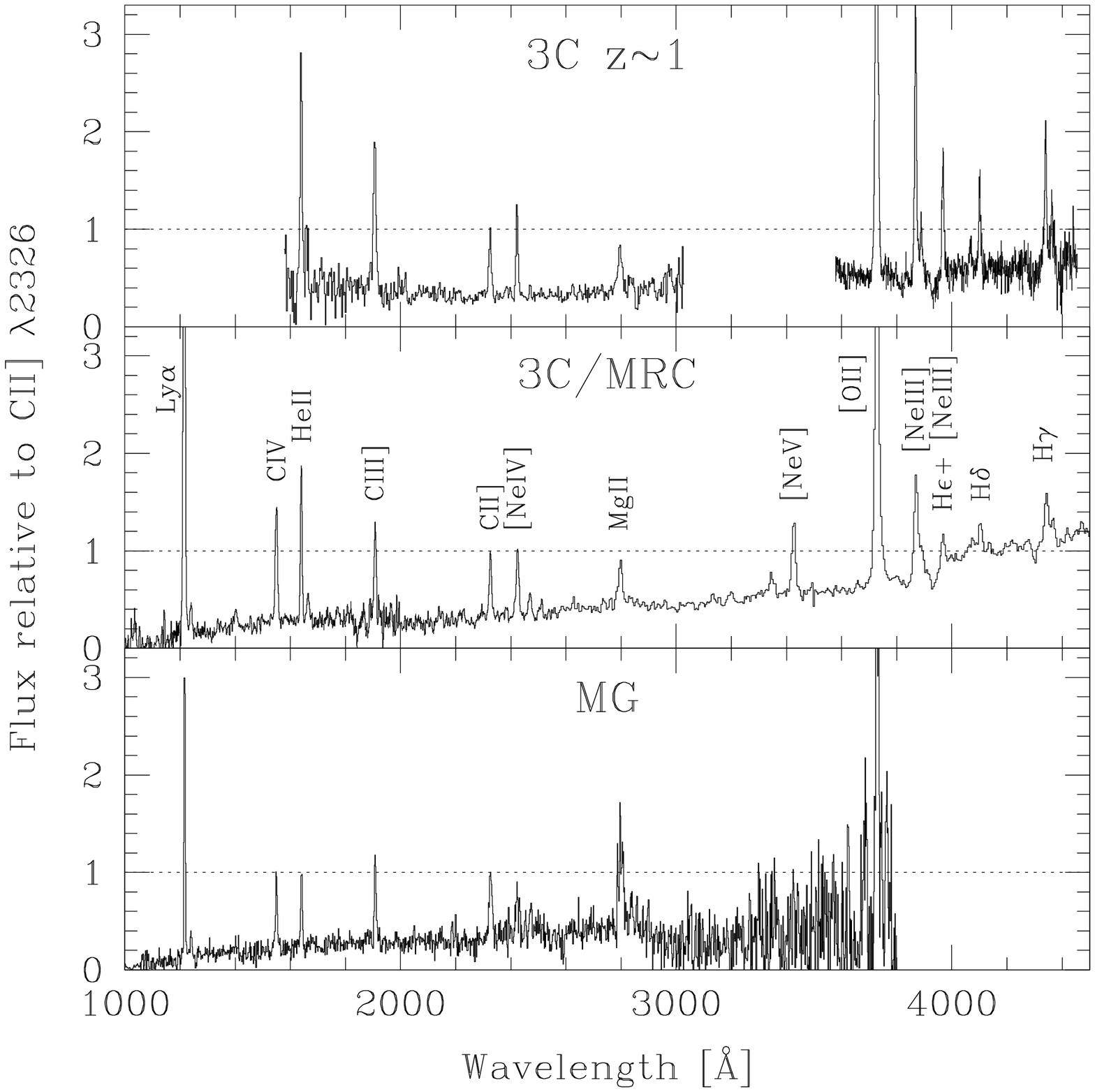,width=8.5cm}
\caption[]{Composite spectra from 3C sources at $z \sim 1$ (\cite{bes00a}), the 3C and MRC samples (\cite{mcc00}) and the MG (\cite{ste99}). All spectra have been normalized to the \CIIfull\ line. The 3C galaxies in the top panel have the largest radio sizes, while the MG galaxies in the bottom panel the have smallest. This is clearly reflected in the dominance of the \CII\ line compared to the \CIV, \HeII\ and \CIII\ lines.}
\label{composites}
\end{figure}
%

The presence of a significant percentage of shock ionized gas in the integrated spectra of HzRGs would explain the poor fit of the pure photo-ionization models to the \CII/\CIII\ ratio (Fig. \ref{CandHediagnostics}). Because of the lack of sources with radio sizes $>$150~kpc, we do not find a population of pure photo-ionization sources like BRL00\nocite{bes00b} do.
The radio size - ionization mechanism relation can also explain the different line ratios in three published composites of HzRGs (Fig. \ref{composites}). The MG sample of S99, which has an angular size cutoff $\theta < 10$\arcsec, contains more sources with relatively strong \CII\ lines. The \CII/\CIII\ ratio in the MG composite spectrum of S99 is 1.6$\times$ stronger than the ratio in the 3C/MRC composite of McCarthy \& Lawrence (2000)\nocite{mcc00}, which does not have an explicit angular size cutoff. This can by easily interpreted if the MG sample contains more sources with important contributions of shock ionization. On the other end, a composite spectrum of 14 3CR galaxies at $0.7<z<1.25$ (\cite{bes00a}) contains more larger radio sources, and should therefore be the most representative composite of the photo-ionization in HzRGs, at least at $z\sim 1$. The \CII/\CIII\ ratio in this composite is 0.3, which is indeed much closer to the pure photo-ionization models.

We can also use Fig. \ref{composites} to identify other emission lines that could be sensitive to shock ionization. The only obvious candidate line is \MgII, which has a constant ratio compared to \CII\ in the 3C($z \sim 1$) and 3C/MRC composites, and is very bright in the MG composite. However, the high luminosity in the MG composite is most likely due to the high selection frequency of the MG survey (5~GHz). At such frequencies, the sources will be more dominated by the flat spectrum radio core, and they will appear more like quasars, which have more prominent \MgII\ lines than radio galaxies, while their \CII\ lines is five times weaker than \MgII\ (see \eg \cite{boy90}). Indeed, S99 showed that the \MgII/\CIII\ ratio in the MG composite is similar to those in quasar composites, and deep spectro-polarimetric observations of radio galaxies and ultra-luminous IR galaxies have revealed a scattered broad \MgII\ component (Tran \etal 1998,2000\nocite{tra98,tra00}).
 
\begin{figure}
\psfig{file=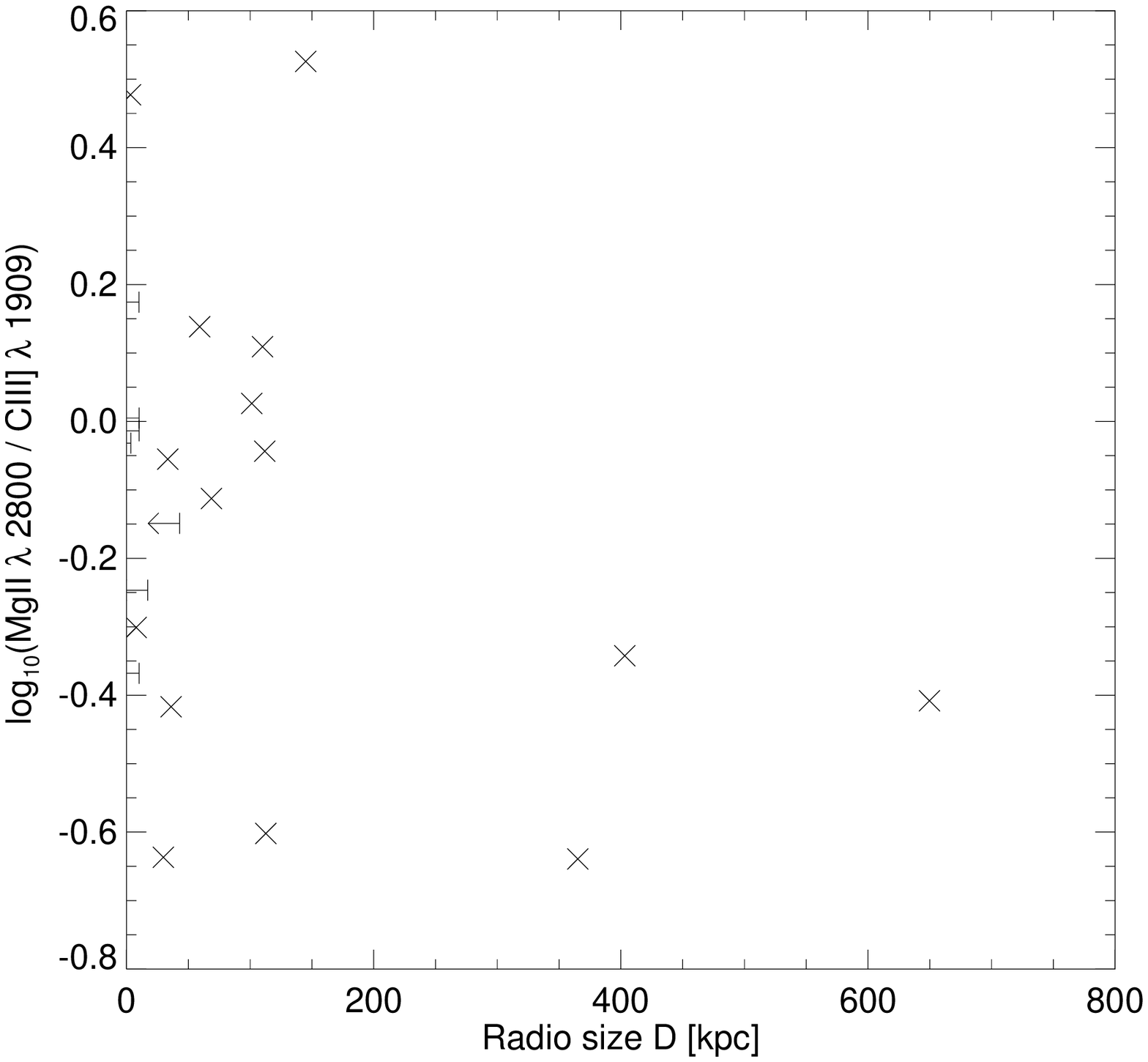,width=8.5cm}
\caption[]{The \MgII/\CIII\ ratio plotted versus radio size. Note the absence of large sources with strong \MgII.}
\label{MgIIoCIIID}
\end{figure}

The plot of the \MgII/\CIII\ ratio versus radio size $D$ in our sample (Fig. \ref{MgIIoCIIID}) looks remarkably similar to the \CII/\CIII\ versus $D$ plot (Fig. \ref{CIIoCIIID}). The highest shock velocity models of DS96 indeed predict \MgII\ to be twice as strong as \CIII, while in a pure photo-ionization model with $n=100$~cm$^{-3}$, $\alpha=-1.0$ and $\log U=-2.25$, \MgII\ would be twice as faint as \CIII. We can therefore also use the \MgII\ line to identify a strong shock contribution to the integrated HzRG spectrum, but should be aware that this line is much more subject to small contributions from obscured broad line regions than \CII.
%

\subsection{Emission-line luminosity - radio power correlations}
In the previous section, we argued that the ionization mechanism in HzRGs is a composite of nuclear photo-ionization and shock ionization dominating the lower ionization lines. Both these processes are likely to be linked to the power of the central engine. To examine this dependence, we shall now consider the relation between the radio power and the emission line luminosity of the shock and photo-ionization sensitive lines. 

In \S4.6.2, we found that the UV line luminosities are weakly correlated with the radio power over $\sim2$ orders of magnitude, with the strongest results for the \CII\ and \Lya\ luminosity. This should be compared with the strong correlation between the \Ha+[NII] or \OII\ luminosity and radio power, which extends over nearly five orders of magnitude in radio power (\eg \cite{bau89}, \cite{raw91}, \cite{mcc93}, \cite{wil99}). This relation has prompted these authors to suggest a common energy source (the AGN) for the {\it total} emission line luminosity and radio power. To calculate the total line luminosity, they have taken the flux in one or two of the brightest lines (\Lya, \OII, \OIII\ or \Ha) and calculated the flux in the other lines using fixed line ratios. 
In HzRGs, the brightest line is \Lya, which can represent up to 1/3 of the total narrow line luminosity (assuming the line ratio from the composite of McCarthy 1993\nocite{mcc93}). We can thus use \Lya\ to calculate the total line luminosity, but this line has the disadvantage that is is highly influenced by resonant scattering and geometrical effects (\eg \cite{vil96}), which will cause the total line flux to be underestimated by a varying amount. 
The weaker correlation with the fainter high ionization lines can be explained by measurement errors and an intrinsic scatter caused by differences in the relation due to different accretion rates, environmental effects, or time variability of the AGN which reaches the narrow-line regions and radio lobes at different times (see Willot \etal 1999 for a detailed discussion).
\begin{figure*}
\psfig{file=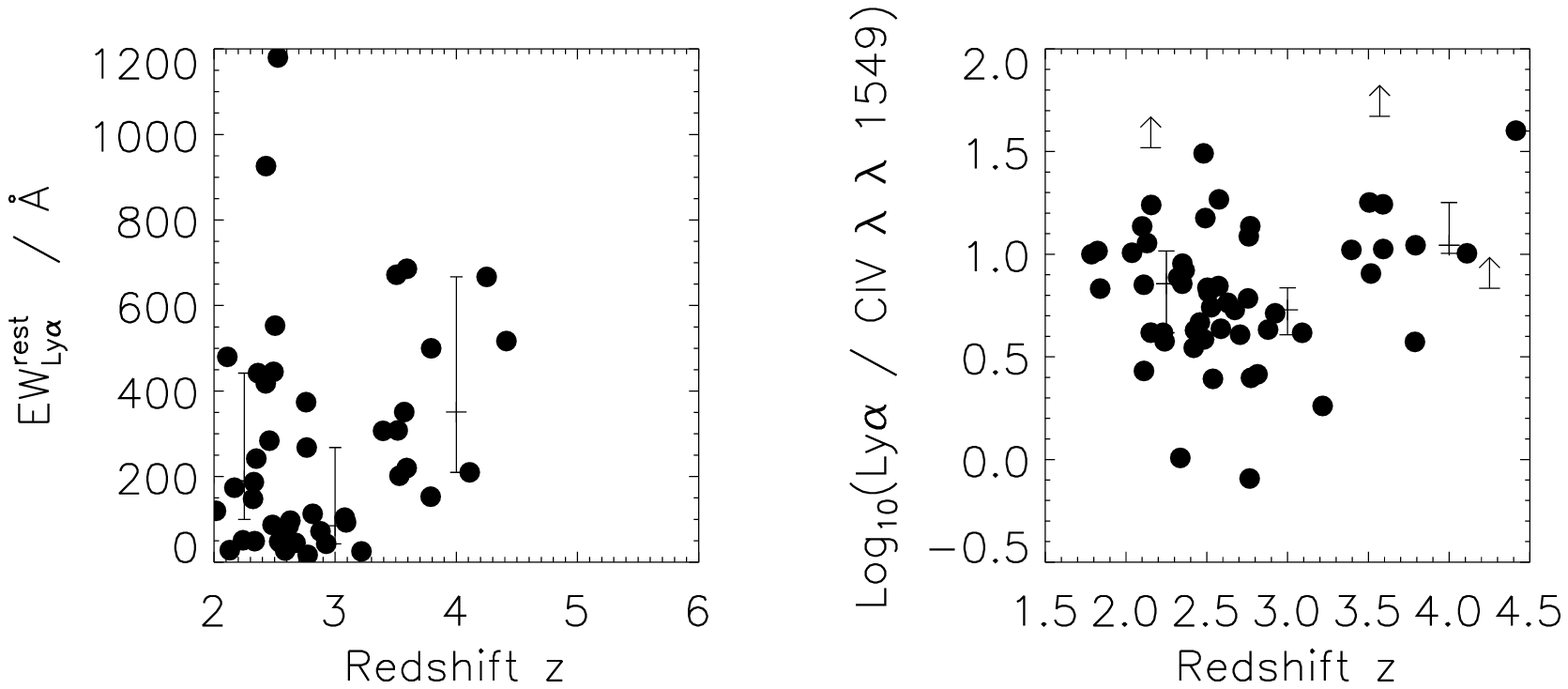,width=16.5cm}
\caption[]{\Lya\ equivalent width (left panel) and \Lya/\CIV\ ratio (right panel) as a function of redshift. Overplotted in each plot are three bins in redshift: $z<2.5$, $2.5<z<3.5$, and $z>3.5$. The tickmarks on each bin are the 25 percentile level, median and 75 percentile levels. Note that the $z>3.5$ bin is always higher than the lower redshift bins, suggesting a higher \Lya\ flux at the highest redshifts.}
\label{Lyaz}
\end{figure*}
%

The fact that the radio power does not correlate with the line ratios (Table \ref{UVratiocorrelations}), but does correlate with the individual UV lines, suggests that in the majority of HzRGs, the total emission line luminosity is dominated by one of the ionization processes (shocks or AGN photo-ionization), and this process has a common energy source with the radio power. Alternatively, it is possible that both mechanisms depend equally strong on radio power.

It is remarkable that the shock ionization dominated \CII\ line luminosity has the strongest correlation with radio power (see Table \ref{radiolineranks}). This suggests that in more luminous radio sources, the shock ionized regions become more prominent. However, the range in radio power in our HzRG sample is still limited, and deeper spectra of less radio-luminous radio galaxies are needed to confirm this trend.

\subsection{The independent behavior of \Lya}
In \S5.4, we found that radio galaxies with relatively over-luminous \Lya\ emission occur more often at $z>3$ than at lower redshifts. The correlation over the entire redshift range $2.0<z<4.4$ is not very strong, and from Table \ref{UVratiocorrelations}, we find that it is not stronger than the correlation with radio size or radio power. However, the rest-frame \Lya\ equivalent width also increases at $z \simgt 3$. Figure \ref{Lyaz} plots the relative intensity of \Lya\ compared with the continuum ($\equiv$ equivalent width; this measure takes some account of absorption by dust), and with \CIV. We find that at $z \simgt 3$, \Lya\ is roughly twice as strong compared to both the continuum and \CIV\ than at $z \simlt 3$. We see similar trends in the ratios with \HeII\ and \CIII, but we have too few data points to make a significant claim. This effect cannot be entirely due to selection effects, as HzRGs with weaker \Lya\ could still have been detected (see Figure \ref{zlinelum}).

We interpret the over-luminous \Lya\ as spectroscopic evidence for the youth of radio galaxies at $z \simgt 3$. At these redshifts, the radio sources found from flux density-limited surveys are inevitably young (\cite{blu99b}). It is therefore conceivable that they are still surrounded by large halos of primordial hydrogen from which the galaxy is formed and which trace the over-dense region that harbours the AGN. The \Lya\ halos at $z \simgt 3$ are often very extended (\eg \cite{oji96}), so we may be witnessing the first epoch of star formation in primordial material. The very frequent occurrence of large amounts of \HI\ absorption in the \Lya\ profiles of the highest redshift radio galaxies (\S4.6.3) also indicates a larger amount of \HI\ surrounding the radio galaxy. As such, the higher hydrogen abundance at $z \simgt 3$ indicates metallicity evolution with redshift in forming massive galaxies. This would be consistent with our findings on the metallicity derived from the \NV/\CIV\ ratio (\S 5.6) and the recent results of Binette \etal (2000)\nocite{bin00}, who found strong evidence from co-spatial \Lya\ and \CIV\ associated absorption that the material in the outer halo must have a much lower metallicity ($Z \sim 0.01 Z_{\odot}$) than the more centrally located emitting material.

At $z\simlt 3$, the radio galaxies become more relaxed, and even the outer parts have experienced chemical enrichment through massive starbursts caused by the infall of the surrounding material on the central (proto-)galaxy, or by the passage of the radio jets. This picture is consistent with the change in rest-frame optical morphology seen in a sample of HzRGs observed in $K-$band by van Breugel \etal (1998)\nocite{wvb98}. At $z \simgt 3$, their morphologies are clumpy and radio-aligned, while at $z \simlt 3$, they often have well organized elliptical profiles, much like their low redshift counterparts.

\section{Conclusions}
We have compiled a sample of 165 HzRG, half of which are at $z>2$. From this sample, we have shown that:

$\bullet$ The different UV line luminosities and equivalent widths are strongly correlated, indicating that the narrow line region gets most of its energy from a common energy source. The UV line luminosities also appear to be correlated with radio power, suggesting a common energy source and either a dominating ionization mechanism or two mechanisms that increase equivalently with radio power. However, the shock ionization sensitive \CII\ line appears to be more strongly correlated than the photo-ionization dominated lines, suggesting the more powerful radio sources can ionize a larger amount of gas by means of shock ionization.

$\bullet$ Using the systematic redshift determined from the \HeII\ line, we introduce a parameter \AL\ measuring the asymmetry in the \Lya\ emission line due to \HI\ absorption. We find that the absorption occurs most frequently at the blue side, and mainly in sources with smaller radio sizes and at higher redshift. The higher \HI\ absorption could trace a denser surrounding medium which can confine the radio source more (\cite{oji97}), or it could reveal an un-pressurized, low density region that has not yet been affected by the passage of a radio jet (\cite{bin00}).

$\bullet$ We confirm the results from Fosbury \etal (1998,1999)\nocite{fos98}\nocite{fos99}, who found that HzRGs occupy a sequence in a \NV/\CIV\ vs. \NV/\HeII\ diagram which is parallel to the metallicity sequence defined by the BLR in quasars. The inclusion of upper limits to the \NV\ flux further extends this sequence to much lower metallicities. The $z>3$ objects occupy the lower metallicity part of this sequence.

$\bullet$ For $z \simgt 3$, \Lya\ becomes over-luminous with respect to both the continuum and the other emission lines. We interpret this as indicating an increasing hydrogen abundance with redshift and evidence that the highest redshift radio galaxies tend to be in an earlier stage of their formation epoch, still surrounded by the reservoir of primordial hydrogen from which the galaxy is forming.

$\bullet$ Pure photo-ionization models with an ionizing continuum of power law spectral index $-1.0$ provide better fits to the line ratios involving \CIV, \HeII, and \CIII\ than shock models, but cannot explain the \CII/\CIII\ ratio. A combination of shock an photo-ionization in the integrated HzRG spectra can better explain the observed ratios, because the \CII\ line is $\sim 5\times$ more sensitive to shocks than photo-ionization. Such a shock ionization component will not be seen in high excitation line ratios, where the photo-ionization will dominate. Diagnostic line-ratio diagrams consisting of only high excitation lines will therefore fail to detect contributions of shock ionization.

$\bullet$ The UV line ratios are not correlated with redshift, radio size, or radio power, with one exception: high \CII/\CIII\ ratios only occur in small radio sources ($\simlt 150$~kpc). This confirms the results of BRL00\nocite{bes00b}, who found that the high \CII/\CIII\ ratios can be explained by shock ionization, but occur only in small radio sources, where the radio size is comparable to the extent of the emission lines regions.
The \CII/\CIII\ ratio in HzRGs requires the addition of a component of shock ionized gas to the integrated HzRG spectra. This is consistent with the generally small radio sizes and high line widths observed. A composite spectrum of HzRGs with small radio sizes (MG) indeed shows a higher contribution of lines sensitive to shock ionization (\CII, \MgII) than composites based on larger radio sources (3C, MRC).

$\bullet$ Within the shock ionization models, the highest shock velocity models provide the best fit to the HzRG spectra. An expansion of these models to higher shock velocities and lower metallicities is needed to determine the full contribution of shock ionization to the integrated HzRG spectra.

\begin{acknowledgements}
We thank Daniel Stern for providing electronic versions of the MG
spectra, and Pat McCarthy for allowing us to use the 3C/MRC composite
spectrum. We also thank Mark Allen, Montse Villar-Mart\'\i n, and
Emmanuel Moy for useful discussions. The work by C.D.B. and W.v.B.\ at
IGPP/LLNL was performed under the auspices of the US Department of
Energy by University of California Lawrence Livermore National
Laboratory under contract W-7405-ENG-48. This work was supported in
part by the Formation and Evolution of Galaxies network set up by the
European Commission under contract ERB FMRX-- CT96--086 of its TMR
programme.

\end{acknowledgements}

\appendix
\section{High redshift radio galaxy sample}
\begin{table*}
\caption[]{High redshift radio galaxy sample}
\label{hzrgsample}
\scriptsize
\begin{tabular}{lrrrrrrrrrrrrr}
\hline
Source & $z$ & LAS$^a$ & Q$^b$ & Log($P_{325}$) & Log($P_{1400}$) & \AL$^c$ & \multicolumn{6}{c}{Line Flux} & Ref. \\
& & \arcsec & & erg/s/Hz & erg/s/Hz & & \multicolumn{6}{c}{$10^{-16}$erg/cm$^2$/s} & \\
 & & & & & & & \Lya & \NV & \CIV & \HeII & \CIII & \CII & \\
\hline
USS 0003$-$019  & 1.541 &    3.3 &\nodata& 35.23 & 34.48 & \nodata&\nodata& \nodata&  5.90 &  3.90 &  3.40 &  1.00 & DB  \\
BRL 0016$-$129  & 1.589 &    3.5 &\nodata& 36.09 & 35.51 & \nodata&\nodata& \nodata&  1.60 &\nodata&  2.60 &  1.20 & BRL \\
MG 0018$+$0940  & 1.586 & $<$1.2 &\nodata& 35.36 & 34.84 & \nodata&\nodata& \nodata&  0.81 &  0.42 &  0.87 &  0.65 & Ste \\
MRC 0030$-$219  & 2.168 & $<$0.3 &\nodata& 35.48 & 34.93 & \nodata&  5.30 & \nodata&\nodata&\nodata&\nodata&\nodata& Mc1 \\
6C 0032$+$412   & 3.670 &    3.0 &\nodata& 35.90 & 35.16 & \nodata&  5.00 & \nodata&\nodata&\nodata&\nodata&\nodata& Lac \\
WN J0040$+$3857 & 2.606 &    1.4 &\nodata& 35.00 & 34.14 &$-$0.13 &  4.10 &$<$0.20 &\nodata&\nodata&\nodata&\nodata& DB  \\
MG 0046$+$1102  & 1.813 & $<$1.2 &\nodata& 35.42 & 34.79 & \nodata&\nodata& \nodata&  0.65 &  0.55 &  0.79 &  0.74 & Ste \\
3C 22.0        & 0.935 &   24.6 &   1.2 & 35.62 & 35.04 & \nodata&\nodata& \nodata&\nodata&\nodata&  6.06 &  1.25 & B00 \\
BRL 0056$-$172  & 1.019 &   17.0 &\nodata& 35.70 & 35.05 & \nodata&\nodata& \nodata&\nodata&\nodata&  1.40 &  0.70 & BRL \\
BRL 0101$-$128  & 0.387 &   16.0 &\nodata& 34.64 & 34.17 & \nodata&\nodata& \nodata&\nodata&\nodata&\nodata&\nodata& BRL \\
TN J0121$+$1320 & 3.516 &    0.3 &\nodata& 35.83 & 34.97 & \nodata&\nodata& \nodata&\nodata&\nodata&\nodata&\nodata& DB  \\
MG 0122$+$1923  & 1.595 & $<$1.2 &\nodata& 35.24 & 34.77 & \nodata&\nodata& \nodata&  0.32 &  0.38 &  0.32 &  0.23 & Ste \\
BRL 0125$-$143  & 0.372 &   15.0 &   1.7 & 34.77 & 34.24 & \nodata&\nodata& \nodata&\nodata&\nodata&\nodata&\nodata& BRL \\
BRL 0128$-$264  & 2.348 &   33.0 &\nodata& 36.48 & 35.81 & \nodata&  7.90 &$<$0.15 &  1.10 &\nodata&  0.60 &\nodata& BRL \\
BRL 0132$+$079  & 0.499 &    9.0 &\nodata& 34.89 & 34.43 & \nodata&\nodata& \nodata&\nodata&\nodata&\nodata&\nodata& BRL \\
TXS J0137$+$2521 & 2.897 &    6.9 &\nodata& 35.88 & 35.16 & \nodata& 12.00 &$<$0.13 &\nodata&\nodata&\nodata&\nodata& DB  \\
MRC 0140$-$257  & 2.616 &    3.4 &\nodata& 35.77 & 35.13 & \nodata&  3.00 & \nodata&\nodata&\nodata&\nodata&\nodata& Mc4 \\
6C 0140$+$326   & 4.413 &    2.6 &   1.1 & 36.03 & 35.30 & \nodata&  6.40 &$<$0.01 &  0.16 &\nodata&\nodata&\nodata& Raw \\
MRC 0152$-$209  & 1.920 &    1.0 &\nodata& 35.61 & 35.03 & \nodata& 18.00 & \nodata&\nodata&\nodata&\nodata&\nodata& Mc4 \\
MRC 0156$-$252  & 2.016 &    8.3 &\nodata& 35.61 & 35.04 & \nodata& 15.00 & \nodata&\nodata&\nodata&\nodata&\nodata& Mc1 \\
USS 0200$+$015  & 2.229 &    5.1 &\nodata& 35.72 & 34.97 &$-$0.04 & 17.40 &$<$0.40 &  4.20 &  3.20 &  4.00 &\nodata& R\"ot \\
TN J0205$+$2242 & 3.506 &    2.7 &\nodata& 35.88 & 35.01 & \nodata&\nodata& \nodata&\nodata&\nodata&\nodata&\nodata& DB  \\
MRC 0203$-$209  & 1.258 &   12.0 &\nodata& 35.06 & 34.47 & \nodata&\nodata& \nodata&\nodata&\nodata&  6.70 &\nodata& Mc1 \\
USS 0211$-$122  & 2.336 &   16.2 &   1.8 & 35.94 & 35.21 &   0.24 &  5.70 &   4.10 &  5.60 &  3.10 &  2.20 &\nodata& R\"ot \\
USS 0214$+$183  & 2.130 &    5.9 &   1.6 & 35.71 & 35.04 & \nodata&\nodata& \nodata&  3.00 &  1.80 &  1.80 &\nodata& R\"ot \\
BRL 0219$+$082  & 0.266 &  155.0 &\nodata& 34.21 & 33.75 & \nodata&\nodata& \nodata&\nodata&\nodata&\nodata&\nodata& BRL \\
WN J0231$+$3600 & 3.079 &   14.8 &\nodata& 35.53 & 34.70 &   0.06 &  1.10 &$<$0.05 &\nodata&\nodata&\nodata&\nodata& DB  \\
WN J0303$+$3733 & 2.504 &    4.4 &\nodata& 35.94 & 35.02 &$-$0.01 & 10.30 &$<$0.20 &  1.50 &\nodata&  1.30 &  0.70 & DB  \\
MG 0311$+$1532  & 1.986 &    5.1 &\nodata& 35.37 & 34.81 & \nodata&\nodata& \nodata&  0.34 &  0.20 &  0.21 &\nodata& Ste \\
BRL 0310$-$150  & 1.769 &$<$10.0 &\nodata& 36.12 & 35.60 & \nodata&\nodata& \nodata& 10.20 &  4.00 &  5.00 &  2.20 & BRL \\
MRC 0316$-$257  & 3.130 &    7.6 &\nodata&\nodata&\nodata& \nodata&  2.40 & \nodata&\nodata&\nodata&\nodata&\nodata& Mc1 \\
MRC 0324$-$228  & 1.894 &    9.6 &\nodata& 35.79 & 35.13 & \nodata& 20.00 & \nodata&\nodata&\nodata&\nodata&\nodata& Mc4 \\
MRC 0349$-$211  & 2.329 &    7.2 &   1.6 & 35.66 & 35.11 & \nodata&  9.00 & \nodata&\nodata&\nodata&\nodata&\nodata& Mc4 \\
USS 0355$-$037  & 2.153 &   11.8 &\nodata& 35.77 & 34.99 & \nodata& 11.20 &$<$0.30 &  2.70 &  3.70 &  2.30 &\nodata& R\"ot \\
BRL 0357$-$163  & 0.584 &    7.0 &\nodata& 35.06 & 34.51 & \nodata&\nodata& \nodata&\nodata&\nodata&\nodata&\nodata& BRL \\
MRC 0406$-$244  & 2.427 &    7.3 &\nodata& 36.46 & 35.62 & \nodata& 47.00 & \nodata&\nodata&\nodata&\nodata&\nodata& Mc4 \\
USS 0417$-$181  & 2.773 &    3.7 &   2.8 & 36.17 & 35.39 &$-$0.14 &  3.00 &$<$0.10 &  1.20 &  0.50 &\nodata&\nodata& R\"ot \\
USS 0448$+$091  & 2.037 &   22.4 &\nodata& 35.31 & 34.64 & \nodata& 12.20 &$<$0.40 &  1.20 &  1.40 &  2.70 &\nodata& R\"ot \\
TN J0452$-$1737 & 2.256 &    2.4 &\nodata& 35.69 & 34.82 & \nodata&\nodata& \nodata&\nodata&  0.30 &  0.10 &  0.40 & DB  \\
4C 60.07       & 3.788 &   16.0 &   2.8 & 36.53 & 35.60 & \nodata& 10.10 &$<$0.10 &  2.70 &\nodata&\nodata&\nodata& R\"ot \\
TN J0516$+$0637 & 0.357 &    1.3 &\nodata& 33.33 & 32.53 & \nodata&\nodata& \nodata&\nodata&\nodata&\nodata&\nodata& DB  \\
BRL 0519$-$208  & 1.086 & $<$2.0 &\nodata& 35.80 & 35.15 & \nodata&\nodata& \nodata&\nodata&\nodata&  6.00 &  0.90 & BRL \\
USS 0529$-$549  & 2.575 &$<$16.0 &\nodata&\nodata&\nodata&$-$0.05 &  7.40 &$<$0.20 &  0.40 &  0.60 &  1.80 &\nodata& R\"ot \\
WN J0617$+$5012 & 3.153 &    3.4 &\nodata& 35.41 & 34.54 & \nodata&  0.80 &$<$0.10 &\nodata&\nodata&\nodata&\nodata& DB  \\
4C 41.17       & 3.792 &   20.0 &   1.2 & 36.45 & 35.66 & \nodata& 14.60 &   0.39 &  1.32 &  0.55 &  0.91 &\nodata& R\"ot \\
B3 0731$+$438   & 2.429 &   10.8 &   1.3 & 36.16 & 35.53 & \nodata& 31.00 & \nodata&\nodata&\nodata&\nodata&\nodata& Mc3 \\
B3 0744$+$464   & 2.926 &    1.9 &   1.4 & 36.19 & 35.56 & \nodata& 12.00 & \nodata&\nodata&\nodata&\nodata&\nodata& Mc3 \\
USS 0748$+$134  & 2.419 &   13.2 &\nodata& 35.72 & 34.97 & \nodata&  6.30 &$<$0.20 &  1.80 &  1.50 &  1.40 &\nodata& R\"ot \\
WN J0813$+$4828 & 1.274 &    0.9 &\nodata& 34.42 & 33.52 & \nodata&\nodata& \nodata&\nodata&\nodata&  0.20 &  0.40 & DB  \\
USS 0828$+$193  & 2.572 & \nodata&   1.2 & 35.55 & 34.84 &$-$0.06 & 13.30 &$<$0.10 &  1.90 &  1.90 &  2.00 &\nodata& R\"ot \\
BRL 0850$-$206  & 1.337 &   13.0 &\nodata& 36.00 & 35.40 & \nodata&\nodata& \nodata&\nodata&\nodata&  2.10 &  0.90 & BRL \\
BRL 0851$-$142  & 1.665 &    7.0 &\nodata& 36.00 & 35.47 & \nodata&\nodata& \nodata&  3.40 &  2.30 &  1.60 &\nodata& BRL \\
4C 12.32      & 2.468 &   15.0 &\nodata&\nodata&\nodata& \nodata& 13.00 & \nodata&\nodata&\nodata&\nodata&\nodata& GK  \\
USS 0857$+$036  & 2.814 &    4.0 &\nodata& 35.87 & 35.16 &$-$0.44 &  2.60 &$<$0.10 &  1.00 &  0.70 &\nodata&\nodata& R\"ot \\
B2 0902$+$34    & 3.395 &    5.0 &   1.2 & 35.97 & 35.41 & \nodata& 21.00 & \nodata&  2.00 &\nodata&\nodata&\nodata& Lil \\
3C 217.0       & 0.898 &   13.1 &   3.3 & 35.61 & 34.99 & \nodata&\nodata& \nodata&\nodata&\nodata&  6.79 &  5.34 & B00 \\
TN J0920$-$0712 & 2.760 &    1.4 &\nodata& 36.00 & 35.05 &$-$0.49 & 44.00 &$<$0.10 &  3.60 &  3.00 &\nodata&\nodata& DB  \\
TN J0924$-$2201 & 5.195 &    1.2 &\nodata& 36.77 & 35.74 & \nodata&  0.35 &$<$0.01 &\nodata&\nodata&\nodata&\nodata& DB  \\
TN J0941$-$1628 & 1.644 &    1.9 &\nodata& 35.71 & 34.87 & \nodata&\nodata& \nodata&  3.20 &  0.90 &  2.00 &  1.90 & DB  \\
3C 226.0       & 0.818 &   31.8 &   1.3 & 35.60 & 34.96 & \nodata&\nodata& \nodata&\nodata&\nodata&  1.94 &  1.00 & B00 \\
USS 0943$-$242  & 2.923 &    3.7 &\nodata& 36.10 & 35.36 &$-$0.37 & 20.10 &   1.70 &  3.90 &  2.70 &  2.30 &\nodata& R\"ot \\
MG 1019$+$0534  & 2.765 &    2.2 &   1.3 & 35.66 & 35.19 &$-$0.68 &  0.84 &   0.23 &  1.04 &  0.85 &  0.49 &\nodata& Ste \\
TN J1033$-$1339 & 2.427 &    2.0 &\nodata& 35.89 & 35.02 &   0.17 &  9.80 &$<$0.10 &  2.30 &  0.80 &  0.70 &\nodata& DB  \\
BRL 1039$+$029  & 0.535 &    6.4 &\nodata& 34.98 & 34.56 & \nodata&\nodata& \nodata&\nodata&\nodata&\nodata&\nodata& BRL \\
3C 247.0       & 0.749 &   13.9 &   1.5 & 35.35 & 34.89 & \nodata&\nodata& \nodata&\nodata&\nodata&  2.73 &  1.18 & B00 \\
TN J1102$-$1651 & 2.111 &    3.0 &\nodata& 35.57 & 34.72 &$-$0.78 &  2.70 &$<$0.20 &  1.00 &  1.30 &  1.10 &\nodata& DB  \\
3C 252.0       & 1.104 &   56.7 &   2.0 & 35.74 & 35.04 & \nodata&\nodata& \nodata&\nodata&  6.16 &  3.56 &  1.38 & B00 \\
TN J1112$-$2948 & 3.090 &    9.1 &   1.5 & 36.00 & 35.12 &   0.22 &  2.90 &$<$0.05 &  0.70 &  1.20 &\nodata&\nodata& DB  \\
USS 1113$-$178  & 2.239 &   10.3 &   2.1 & 35.72 & 35.08 & \nodata&  6.40 &$<$0.20 &  1.70 &  0.70 &  2.80 &\nodata& R\"ot \\
3C 256.0       & 1.824 &    4.0 &   2.9 & 36.15 & 35.51 & \nodata& 54.20 &   1.40 &  5.23 &  5.47 &  4.28 &  2.03 & Sim \\
WN J1123$+$3141 & 3.217 &   25.8 &   1.6 & 35.99 & 35.06 &$-$0.18 &  6.20 &$<$1.00 &  3.40 &  2.40 &\nodata&\nodata& DB  \\
USS 1138$-$262  & 2.156 &   11.4 &   1.5 & 36.34 & 35.57 & \nodata& 13.90 &$<$0.30 &  0.80 &  1.30 &  1.30 &\nodata& R\"ot \\
BRL 1138$+$015  & 0.443 &    5.2 &\nodata& 34.80 & 34.39 & \nodata&\nodata& \nodata&\nodata&\nodata&\nodata&\nodata& BRL \\
MG 1142$+$1338  & 1.279 & $<$0.4 &\nodata& 34.88 & 34.50 & \nodata&\nodata& \nodata&\nodata&\nodata&\nodata&\nodata& Ste \\
BRL 1140$-$114  & 1.935 &    3.9 &\nodata& 36.30 & 35.60 & \nodata&\nodata& \nodata&  1.00 &  0.50 &  0.60 &  1.90 & BRL \\
B2 1141$+$354   & 1.781 &   11.0 &\nodata& 35.45 & 34.82 & \nodata& 47.00 & \nodata&\nodata&\nodata&  0.80 &\nodata& AS  \\
3C 265.0       & 0.810 &   78.8 &   1.5 & 35.80 & 35.07 & \nodata&\nodata& \nodata&\nodata&\nodata& 22.78 &  3.56 & B00 \\
4C 26.38       & 2.608 &   23.0 &\nodata&\nodata&\nodata& \nodata&\nodata& \nodata&  8.90 &  5.70 &  2.40 &\nodata& R\"ot \\
B2 1230$+$349   & 1.533 &   11.0 &\nodata& 35.44 & 34.85 & \nodata&\nodata& \nodata&\nodata&\nodata&  3.00 &  2.00 & AS  \\
6C 1232$+$39    & 3.220 &    8.0 &\nodata& 36.28 & 35.48 & \nodata& 10.00 & \nodata&\nodata&\nodata&\nodata&\nodata& Lac \\
VLA~J123642$+$621331  & 4.424 &    0.2 &\nodata&\nodata&\nodata& \nodata&  0.07 & \nodata&\nodata&\nodata&\nodata&\nodata& Wad \\
USS 1243$+$036  & 3.570 &    7.0 &   1.4 &\nodata&\nodata& \nodata& 23.50 &$<$0.50 &\nodata&\nodata&\nodata&\nodata& R\"ot \\
MG 1251$+$1104  & 2.322 & $<$1.2 &\nodata&\nodata&\nodata& \nodata&  2.31 &$<$0.06 &  0.30 &  0.30 &  0.52 &\nodata& Ste \\
3C 280.0       & 0.997 &   13.7 &   1.2 & 35.94 & 35.43 & \nodata&\nodata& \nodata&\nodata& 10.76 &  7.17 &  1.50 & B00 \\
BRL 1303$+$091  & 1.409 &    8.0 &   2.9 & 35.96 & 35.30 & \nodata&\nodata& \nodata&\nodata&  2.60 &  3.50 &\nodata& BRL \\
BRL 1307$+$000  & 0.419 &   60.0 &   1.1 & 34.61 & 34.13 & \nodata&\nodata& \nodata&\nodata&\nodata&\nodata&\nodata& BRL \\
\end{tabular}
\end{table*}

\begin{table*}
\scriptsize
\begin{tabular}{lrrrrrrrrrrrrr}
\hline
Source & $z$ & LAS$^a$ & Q$^b$ & Log($P_{325}$) & Log($P_{1400}$) & \AL$^c$ & \multicolumn{6}{c}{Line Flux} & Ref. \\
& & \arcsec & & erg/s/Hz & erg/s/Hz & & \multicolumn{6}{c}{$10^{-16}$erg/cm$^2$/s} & \\
 & & & & & & & \Lya & \NV & \CIV & \HeII & \CIII & \CII & \\
\hline
WN J1333$+$3037 & 1.213 &    0.4 &\nodata& 34.94 & 33.65 & \nodata&\nodata& \nodata&\nodata&\nodata&  0.30 &  0.40 & DB  \\
WN J1338$+$3532 & 2.769 &   11.6 &\nodata& 35.72 & 34.97 &$-$0.01 & 17.80 &$<$0.60 &  1.30 &  3.00 &  2.20 &\nodata& DB  \\
TN J1338$-$1942 & 4.110 &    1.4 &   2.7 & 36.29 & 35.46 &$-$0.50 & 10.10 &$<$0.07 &  1.00 &  0.20 &\nodata&\nodata& DB  \\
3C 289.0       & 0.967 &   10.6 &   1.1 & 35.65 & 35.09 & \nodata&\nodata& \nodata&\nodata&\nodata&  2.81 &  1.01 & B00 \\
BRL 1344$-$078  & 0.384 &$<$10.0 &\nodata& 34.72 & 34.15 & \nodata&\nodata& \nodata&\nodata&\nodata&\nodata&\nodata& BRL \\
4C 24.28       & 2.879 &    7.0 &   1.1 & 36.39 & 35.65 & \nodata&  7.30 &   6.90 &  1.70 &\nodata&\nodata&\nodata& R\"ot \\
WN J1356$+$3929 & 0.253 &    1.6 &\nodata& 33.67 & 32.86 & \nodata&\nodata& \nodata&\nodata&\nodata&\nodata&\nodata& DB  \\
BRL 1354$+$013  & 0.819 &   33.0 &\nodata&\nodata&\nodata& \nodata&\nodata& \nodata&\nodata&\nodata&\nodata&\nodata& BRL \\
USS 1357$+$007  & 2.673 &    4.6 &\nodata&\nodata&\nodata&$-$0.06 &  9.10 &$<$0.20 &  1.70 &\nodata&\nodata&\nodata& R\"ot \\
MG 1401$+$0921  & 2.093 &    3.5 &\nodata& 35.53 & 34.92 & \nodata&\nodata& \nodata&  0.41 &  0.50 &  0.34 &  0.17 & Ste \\
TN J1402$-$1510 & 0.739 &   15.9 &\nodata& 35.00 & 34.38 & \nodata&\nodata& \nodata&\nodata&\nodata&\nodata&\nodata& DB  \\
3C 294.0       & 1.786 &   14.5 &   1.4 &\nodata&\nodata& \nodata&   155 &   3.10 & 15.50 & 15.50 & 18.60 &\nodata& Mc2 \\
USS 1410$-$001  & 2.363 &   23.5 &   1.0 & 35.65 & 35.00 &   0.06 & 28.00 &   1.68 &  3.36 &  2.52 &  1.12 &  0.36 & Cim \\
BRL 1411$-$057  & 1.094 &   47.0 &   1.4 & 35.72 & 35.06 & \nodata&\nodata& \nodata&\nodata&\nodata&  1.10 &\nodata& BRL \\
BRL 1422$-$297  & 1.632 &$<$10.0 &\nodata& 36.11 & 35.57 & \nodata&\nodata& \nodata&  4.30 &  2.10 &  1.00 &\nodata& BRL \\
USS 1425$-$148  & 2.349 &   10.7 &\nodata& 35.88 & 35.28 &$-$0.01 & 20.70 &$<$0.70 &  2.30 &  2.30 &  1.00 &\nodata& DB  \\
8C 1435$+$635   & 4.250 &    4.3 &   1.1 & 36.95 & 36.12 & \nodata&  1.50 & \nodata&\nodata&\nodata&\nodata&\nodata& Spi \\
USS 1436$+$157  & 2.538 &    4.7 &   1.5 &\nodata&\nodata&$-$0.01 & 42.00 &$<$0.80 & 17.00 &  6.00 &  9.40 &\nodata& R\"ot \\
BRL 1436$-$167  & 0.146 &$<$12.0 &\nodata& 33.77 & 33.29 & \nodata&\nodata& \nodata&\nodata&\nodata&\nodata&\nodata& BRL \\
BRL 1509$+$015  & 0.792 &    7.2 &   1.4 & 35.37 & 34.87 & \nodata&\nodata& \nodata&\nodata&\nodata&\nodata&\nodata& BRL \\
USS 1545$-$234  & 2.755 &    6.3 &   1.9 & 35.92 & 35.24 &   0.01 &  6.70 &$<$0.20 &  1.10 &  1.80 &\nodata&\nodata& R\"ot \\
3C 324.0       & 1.208 &   11.1 &   1.7 & 35.92 & 35.34 & \nodata&\nodata& \nodata&  3.67 &  2.70 &  3.47 &  1.43 & B00 \\
USS 1558$-$003  & 2.527 &    7.7 &\nodata&\nodata&\nodata&   0.18 & 14.90 &$<$0.10 &  2.70 &  1.70 &  1.20 &\nodata& R\"ot \\
BRL 1602$-$174  & 2.043 &   37.0 &   3.2 & 36.32 & 35.67 & \nodata&\nodata& \nodata& 10.00 &  4.80 &  2.70 &\nodata& BRL \\
BRL 1602$-$288  & 0.482 &   61.0 &   1.0 & 35.09 & 34.48 & \nodata&\nodata& \nodata&\nodata&\nodata&\nodata&\nodata& BRL \\
BRL 1602$-$093  & 0.109 &  290.0 &\nodata&\nodata&\nodata& \nodata&\nodata& \nodata&\nodata&\nodata&\nodata&\nodata& BRL \\
BRL 1603$+$001  & 0.059 &   11.0 &\nodata& 32.96 & 32.22 & \nodata&\nodata& \nodata&\nodata&\nodata&\nodata&\nodata& BRL \\
BRL 1621$-$115  & 0.375 &$<$20.0 &\nodata& 34.73 & 34.25 & \nodata&\nodata& \nodata&\nodata&\nodata&\nodata&\nodata& BRL \\
3C 340.0       & 0.775 &   44.7 &   1.1 & 35.47 & 34.84 & \nodata&\nodata& \nodata&\nodata&\nodata&  7.08 &  0.75 & B00 \\
BRL 1628$-$268  & 0.166 &   93.0 &\nodata&\nodata&\nodata& \nodata&\nodata& \nodata&\nodata&\nodata&\nodata&\nodata& BRL \\
BRL 1643$+$022  & 0.095 &    7.1 &   1.3 &\nodata&\nodata& \nodata&\nodata& \nodata&\nodata&\nodata&\nodata&\nodata& BRL \\
TXS J1650$+$0955 & 2.510 &   19.4 &\nodata&\nodata&\nodata&   0.28 & 20.90 &$<$0.10 &  3.20 &  2.70 &  1.20 &\nodata& DB  \\
BRL 1649$-$062  & 0.236 &   85.0 &\nodata&\nodata&\nodata& \nodata&\nodata& \nodata&\nodata&\nodata&\nodata&\nodata& BRL \\
WN J1703$+$3739 & 0.256 &   30.1 &\nodata& 31.98 & 31.41 & \nodata&\nodata& \nodata&\nodata&\nodata&\nodata&\nodata& DB  \\
USS 1707$+$105  & 2.349 &   21.7 &\nodata&\nodata&\nodata& \nodata&  4.30 &$<$0.10 &\nodata&  0.90 &  0.20 &\nodata& R\"ot \\
3C 352.0       & 0.806 &   12.3 &   1.6 & 35.50 & 34.84 & \nodata&\nodata& \nodata&\nodata&\nodata&  4.11 &  2.36 & B00 \\
3C 356.0       & 1.079 &   73.4 &\nodata& 35.72 & 35.04 & \nodata&\nodata& \nodata&\nodata&  4.33 &  4.41 &  0.52 & B00 \\
BRL 1732$-$092  & 0.317 &   45.0 &\nodata& 34.44 & 34.00 & \nodata&\nodata& \nodata&\nodata&\nodata&\nodata&\nodata& BRL \\
8C 1736$+$650   & 2.400 &   17.0 &\nodata& 35.08 & 34.24 & \nodata& 18.00 & \nodata&\nodata&\nodata&\nodata&\nodata& L99 \\
7C 1740$+$6640  & 2.100 & $<$0.5 &\nodata& 34.84 & 34.50 & \nodata&  4.10 & \nodata&  0.30 &\nodata&\nodata&\nodata& L99 \\
7C 1758$+$6719  & 2.700 &   45.0 &\nodata& 35.40 & 34.69 & \nodata& 12.00 & \nodata&\nodata&\nodata&  0.80 &\nodata& L99 \\
7C 1802$+$6456  & 2.110 &   26.0 &\nodata& 35.54 & 34.76 & \nodata&  7.10 & \nodata&  1.00 &\nodata&  0.80 &\nodata& L99 \\
8C 1803$+$661   & 1.610 &   36.0 &\nodata& 34.55 & 33.93 & \nodata&\nodata& \nodata&  5.30 &  2.60 &  1.90 &\nodata& L99 \\
3C 368.0       & 1.132 &    8.5 &   2.0 & 35.93 & 35.10 & \nodata&\nodata& \nodata&\nodata&  5.11 & 10.57 &  9.98 & B00 \\
7C 1805$+$6332  & 1.840 &   14.0 &\nodata& 35.13 & 34.47 & \nodata& 17.00 & \nodata&  2.50 &\nodata&\nodata&\nodata& L99 \\
7C 1807$+$6719  & 2.780 &    1.9 &\nodata& 35.31 & 34.78 & \nodata&  2.10 &   0.70 &\nodata&\nodata&\nodata&\nodata& L99 \\
4C 40.36       & 2.265 &   20.0 &\nodata& 36.24 & 35.42 & \nodata&\nodata& \nodata&  6.20 &  5.60 &  5.90 &\nodata& R\"ot \\
BRL 1859$-$235  & 1.430 &    4.2 &\nodata& 36.29 & 35.68 & \nodata&\nodata& \nodata&  3.40 &  4.60 &  4.70 &  6.70 & BRL \\
TXS J1908$+$7220 & 3.530 &   14.4 &\nodata& 36.44 & 35.64 & \nodata& 32.00 &$<$0.20 &\nodata&\nodata&\nodata&\nodata& DB  \\
WN J1911$+$6342 & 3.590 &    1.8 &\nodata& 35.56 & 34.66 &$-$0.64 &  1.40 &$<$0.03 &  0.08 &  0.08 &\nodata&\nodata& DB  \\
BRL 1912$-$269  & 0.226 &   48.0 &   1.1 & 34.33 & 33.57 & \nodata&\nodata& \nodata&\nodata&\nodata&\nodata&\nodata& BRL \\
BRL 1920$-$077  & 0.648 &   23.0 &\nodata& 35.19 & 34.62 & \nodata&\nodata& \nodata&\nodata&\nodata&\nodata&\nodata& BRL \\
4C 48.48       & 2.343 &   17.0 &   1.1 & 35.87 & 35.20 & \nodata&\nodata& \nodata&  6.10 &  3.70 &  2.80 &\nodata& R\"ot \\
MRC 2025$-$218  & 2.630 &    5.1 &\nodata& 35.89 & 35.25 & \nodata&  4.00 &   0.62 &  0.69 &  0.35 &  0.97 &\nodata& VM99 \\
TXS J2036$+$0256 & 2.130 &    3.1 &\nodata& 35.86 & 35.09 &   0.01 &  6.80 &$<$0.30 &  0.60 &  0.70 &  1.20 &\nodata& DB  \\
MG 2037$-$0011  & 1.512 & $<$0.4 &\nodata& 34.93 & 34.63 & \nodata&\nodata& \nodata&\nodata&\nodata&  0.44 &  0.10 & Ste \\
MG 2058$+$0542  & 1.381 & $<$0.4 &\nodata& 35.40 & 35.02 & \nodata&\nodata& \nodata&\nodata&\nodata&  0.71 &  0.41 & Ste \\
MRC 2104$-$242  & 2.491 &   21.8 &\nodata& 36.26 & 35.41 & \nodata& 57.00 &$<$3.80 &  3.80 &  1.90 &  2.66 &\nodata& VM99 \\
4C 23.56       & 2.483 &   47.0 &   1.9 &\nodata&\nodata& \nodata&  8.00 &   1.36 &  2.08 &  1.52 &  1.28 &  0.80 & Cim \\
MG 2109$+$0326  & 1.634 & $<$1.2 &\nodata& 35.28 & 34.71 & \nodata&\nodata& \nodata&\nodata&  0.32 &  0.21 &  0.23 & Ste \\
MRC 2115$-$253  & 1.114 &    1.7 &\nodata& 35.16 & 34.45 & \nodata&\nodata& \nodata&\nodata&\nodata& 37.00 &\nodata& Mc1 \\
MG 2121$+$1839  & 1.860 &    6.3 &\nodata& 35.32 & 34.73 & \nodata&\nodata& \nodata&  0.53 &  0.14 &  0.24 &  0.15 & Ste \\
BRL 2120$-$166  & 0.882 &   14.0 &\nodata& 35.49 & 34.86 & \nodata&\nodata& \nodata&\nodata&\nodata&\nodata&  1.60 & BRL \\
MG 2144$+$1928  & 3.592 &    8.5 &\nodata& 36.37 & 35.65 &$-$0.47 &  6.15 &$<$0.04 &  0.58 &  0.35 &\nodata&\nodata& Ste \\
USS 2202$+$128  & 2.706 &    3.4 &   1.4 & 35.79 & 35.10 & \nodata&  7.70 &$<$0.10 &  1.90 &  2.40 &\nodata&\nodata& R\"ot \\
3C 441.0       & 0.708 &   26.0 &   3.1 & 35.38 & 34.82 & \nodata&\nodata& \nodata&\nodata&\nodata&\nodata&  0.30 & B00 \\
MRC 2224$-$273  & 1.679 &    0.4 &\nodata& 35.37 & 34.72 & \nodata& 47.00 & \nodata&\nodata&\nodata&\nodata&\nodata& Mc4 \\
MRC 2247$-$232  & 1.326 &    9.3 &\nodata& 35.57 & 35.01 & \nodata&\nodata& \nodata&\nodata&\nodata&  1.50 &\nodata& Mc4 \\
USS 2251$-$089  & 1.986 &    4.5 &   1.4 & 35.66 & 35.07 & \nodata&\nodata& \nodata&  3.30 &  1.30 &  1.50 &\nodata& R\"ot \\
TN J2254$+$1857 & 2.153 &    2.7 &\nodata& 34.97 & 34.15 & \nodata& 33.00 &$<$1.00 &\nodata&  3.20 &\nodata&\nodata& DB  \\
MRC 2303$-$253  & 0.740 &   18.6 &\nodata& 35.05 & 34.35 & \nodata&\nodata& \nodata&\nodata&\nodata&  7.00 &\nodata& Mc1 \\
MG 2308$+$0336  & 2.457 & \nodata&\nodata& 35.89 & 35.32 &$-$0.74 &  2.93 &   0.57 &  0.63 &  0.39 &  0.45 &  0.83 & Ste \\
BRL 2318$-$166  & 1.414 & $<$5.0 &\nodata& 36.20 & 35.52 & \nodata&\nodata& \nodata&\nodata&  7.80 &  6.20 &  2.00 & BRL \\
MRC 2318$-$244  & 1.113 &   25.0 &   1.1 & 35.29 & 34.74 & \nodata&\nodata& \nodata&\nodata&\nodata&  7.00 &\nodata& Mc4 \\
TXS J2321$+$2237 & 2.553 &    8.1 &\nodata& 35.90 & 35.14 &   0.17 &  8.10 &$<$0.20 &\nodata&\nodata&\nodata&\nodata& DB  \\
BRL 2322$-$052  & 1.188 &    7.8 &\nodata&\nodata&\nodata& \nodata&\nodata& \nodata&\nodata&\nodata&\nodata&  1.80 & BRL \\
TXS J2334$+$1545 & 2.480 &    6.4 &\nodata& 35.69 & 34.98 & \nodata&  3.10 &$<$0.10 &  0.10 &\nodata&  0.80 &\nodata& DB  \\
BRL 2347$-$026  & 1.036 & $<$2.0 &\nodata& 35.56 & 35.03 & \nodata&\nodata& \nodata&\nodata&\nodata&\nodata&  0.60 & BRL \\
TXS J2351$+$1034 & 1.334 &    2.2 &\nodata& 35.39 & 34.60 & \nodata&\nodata& \nodata&\nodata&\nodata&\nodata&\nodata& DB  \\
4C 28.58       & 2.891 &   28.0 &   2.3 & 36.30 & 35.44 & \nodata&\nodata& \nodata&  0.30 &  1.60 &  1.80 &\nodata& R\"ot \\
TXS J2355$-$0002 & 2.587 &   33.8 &   1.6 & 35.96 & 35.28 & \nodata&  1.30 &   0.80 &  0.30 &\nodata&\nodata&\nodata& DB  \\
\hline 
\end{tabular}

$^a$ Radio largest angular size, see \S3.1.2

$^b$ Radio lobe distance ratio, see \S3.1.3

$^c$ \Lya\ asymmetry, see \S3.2.2

REFERENCES: AS=\cite{ali88}; BRL=\cite{bes99}; B00=\cite{bes00a}; Cim=\cite{cim98b}; DB=\cite{deb00b}; GK=\cite{gop95}; Lac=\cite{lac94}; L99=\cite{lac99}; Lil=\cite{lil88}; Mc1=\cite{mcc90a}; Mc2=\cite{mcc90a}; Mc3=\cite{mcc91b}; Mc4=\cite{mcc91c}; Raw=\cite{raw96}; R\"ot=\cite{rot97}; Sim=\cite{sim99}; Spi=\cite{spi95}; Ste=\cite{ste99}; VM99=\cite{vil99c}; Wad=\cite{wad99}
\end{table*}

\section{Spearman rank correlation coefficients}
\begin{figure*}
\psfig{file=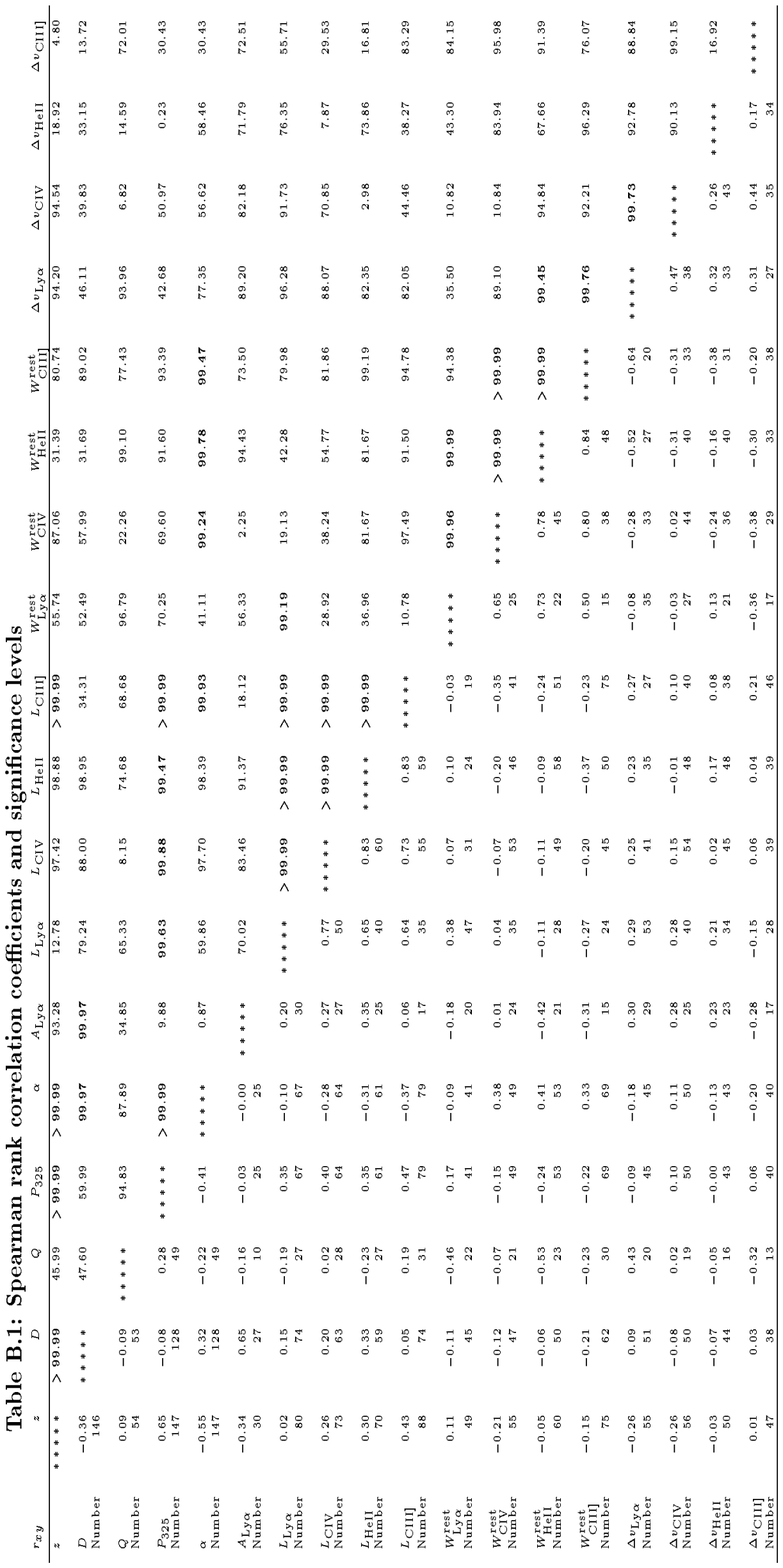,angle=180,width=20cm}
\end{figure*}


\begin{thebibliography}{}

\bibitem[Adam \etal 1997]{ada97} Adam, G., Rocca-Volmerange, B., G\'erard, S., Ferruit, P., \& Bacon, R. 1997, \aap, 326, 501

\bibitem[Allen, Dopita \& Tsvetanov 1998]{all98} Allen, M., Dopita, M, \& Tsvetanov, Z. 1998, \apj, 493, 571 (ADT98)

\bibitem[Allington-Smith 1982]{ali82} Allington-Smith, J. 1982, \mnras, 199, 611 

\bibitem[Allington-Smith \etal 1988]{ali88} Allington-Smith, J., Spinrad, H., Djorgovski, S., \& Liebert, J. 1988, \mnras, 234, 1091

\bibitem[Archibald \etal 2000]{arc00} Archibald, E., Dunlop, J., Hughes, D., Rawlings, S., Eales, S., \& Ivison, R. 2000, \mnras, in press, astro-ph/0002083

\bibitem[Baum \& Heckman 1989]{bau89} Baum, S. \& Heckman, T. 1989, \apj, 336, 702

\bibitem[Baum \& McCarthy 2000]{bau00} Baum, S. \& McCarthy, P. 2000, \aj, in press, astro-ph/0002329

\bibitem[Best \etal 1995]{bes95} Best, P., Bailer, D., Longair, M. \& Riley, J. 1995, \mnras, 275, 1171 

\bibitem[Best, Longair \& R\"ottgering 1996]{bes96} Best, P., Longair, M. \& R\"ottgering, H. 1996, \mnras, 280, L9

\bibitem[Best, Longair \& R\"ottgering 1998]{bes98} Best, P., Longair, M., \& R\"ottgering, H. 1998, \mnras, 295, 549

\bibitem[Best, R\"ottgering \& Lehnert 1999]{bes99} Best, P., R\"ottgering, H. \& Lehnert, M. 1999, \mnras, 310, 223 

\bibitem[Best, R\"ottgering \& Longair 2000a]{bes00a} Best, P., R\"ottgering, H. \& Longair, M. 2000a, \mnras, 311, 1 

\bibitem[Best, R\"ottgering \& Longair 2000b]{bes00b} Best, P., R\"ottgering, H. \& Longair, M. 2000b, \mnras, 311, 23 (BRL00)

\bibitem[Bicknell \etal 2000]{bic00} Bicknell, G., Sutherland, R., van Breugel, W., Dopita, M., Dey, A., \& Miley, G. 2000, \apj, in press, astro-ph/9909218

\bibitem[Binette, Wilson \& Storchi-Bergmann 1996]{bin96} Binette, L., Wilson, A. \& Storchi-Bergmann, T. 1996, \aap, 312, 365 (BWS96)

\bibitem[Binette \etal 2000]{bin00} Binette, L., Kurk, J., Villar-Mart\'\i n, M \& R\"ottgering, H. 2000, \aap, in press, astro-ph/0002210

\bibitem[Blundell \etal\ 1998]{blu98} Blundell, K., Rawlings, S., Eales, S., Taylor, G., \& Bradley, A. 1998, \mnras, 295, 265

\bibitem[Blundell, Rawlings \& Willot 1999]{blu99a} Blundell, K., Rawlings, S., \& Willot, C. 1999, \aj, 117, 677

\bibitem[Blundell \& Rawlings 1999]{blu99b} Blundell, K. \& Rawlings, S. 1999, \nature, 399, 330 

\bibitem[Boyle 1990]{boy90} Boyle, B. J. 1990, \mnras, 243, 231 

\bibitem[Carilli \etal\ 1997]{car97} Carilli, C., R\"ottgering, H., van Ojik, R., Miley, G., \& van Breugel, W. 1997, \apjs, 109, 1

\bibitem[Chambers, Miley \& van Breugel 1987]{cha87} Chambers, K., Miley, G., \& van Breugel, W. 1987, \nature, 329, 604 

\bibitem[Chambers, Miley \& van Breugel 1990]{cha90} Chambers, K., Miley, G., \& van Breugel, W. 1990, \apj, 363, 21

\bibitem[Chini \& Kr\"ugel 1994]{chi94} Chini, R., \& Kr\"ugel, E. 1994, \aap, 288, L33 

\bibitem[Cimatti \etal 1998a]{cim98a} Cimatti, A., Freudling, W., R\"ottgering, H., Ivison, R., \& Mazzei, P. 1998, \aap, 329, 399

\bibitem[Cimatti \etal 1998b]{cim98b} Cimatti, A., di Serego Alighieri, S., Vernet, J., Cohen, M., \& Fosbury, R. 1998, \apjl, 499, L21 

\bibitem[Condon \etal 1998]{con98} Condon, J. J. \etal 1998, \aj, 115, 1693

\bibitem[De Breuck \etal 1999]{deb99} De Breuck, C., van Breugel, W., Minniti, D.,  Miley, G., R\"ottgering, H., \& Carilli, C. 1999, \aap, 352, L51

\bibitem[De Breuck \etal 2000a]{deb00a} De Breuck, C., van Breugel, W., R\"ottgering, H., \& Miley, G. 2000a, \aasup, 143, 303

\bibitem[De Breuck \etal 2000b]{deb00b} De Breuck, C., van Breugel, W., R\"ottgering, H., Miley, G., \& Stern, D., 2000b, \apj, in preparation

\bibitem[Dey, Spinrad \& Dickinson 1995]{dey95} Dey, A., Spinrad, H., \& Dickinson, M. 1995, \apj, 440, 515 

\bibitem[Dey \etal 1997]{dey97} Dey, A., van Breugel, W., Vacca, W., \& Antonucci, R. 1997, \apj, 449, 698

\bibitem[Dey et al. 1998]{dey98} Dey, A., Spinrad, H., Stern, D., Graham, J. R., \& Chaffee, F. H. 1998, \apjl, 498, L93 

\bibitem[Dey 1999]{dey99} Dey, A. 1999, in ``The Most Distant Radio Galaxies'', ed. H. R\"ottgering, P. Best \ M. Lehnert (Amsterdam: KNAW), p. 19

\bibitem[Dopita \& Sutherland 1996]{dop96} Dopita, M., \& Sutherland, R. 1996, \apjs, 102, 161 (DS96)

\bibitem[Douglas \etal\ 1996]{dou96} Douglas, J., Bash, F., Bozyan, F., Torrence, G., \& Wolfe, C. 1996, \aj, 111, 1945

\bibitem[Dunlop \& Peacock 1990]{dun90} Dunlop, J., \& Peacock, J. 1990, \mnras, 247, 19 

\bibitem[Dunlop \& 1994]{dun94} Dunlop, J., Hughes, D., Rawlings, S., Eales, S., \& Ward, M. 1994, \nature, 370, 347 

\bibitem[Eales \& Rawlings 1993]{eal93} Eales, S., \& Rawlings, S. 1993, \apj, 411, 67 

\bibitem[Eales \etal 1997]{eal97} Eales, S., Rawlings, S., Law-Green, D., Cotter, G., \& Lacy, M.\ 1997, \mnras, 291, 593

\bibitem[Evans 1998]{eva98} Evans, A. 1998, \apj, 498, 553 

\bibitem[Evans \etal 1999]{eva99} Evans, I., Koratkar, A., Allen, M., Dopita, M., \& Tsvetanov, Z. 1999, \apj, 521, 531

\bibitem[Ferland 1996]{fer96} Ferland, G. 1996, Hazy: A Brief Introduction to CLOUDY, Univ. of Kentucky Dept. of Phys. and Astron. Internal Report

\bibitem[Fosbury \etal 1998]{fos98} Fosbury, R. \etal 1998, in NICMOS and the VLT: A New Era of High Resolution Near Infrared Imaging and Spectroscopy, ESO Conference and Workshop Proceedings 55, W. Freudling and R. Hook eds., p. 190

\bibitem[Fosbury \etal 1999]{fos99} Fosbury, R. \etal 1999, in ESO conference on Chemical Evolution from Zero to High Redshift, ESO astrophysics symposia, J. Walsh \& M. Rosa eds., in press, astro-ph/9901115

\bibitem[Gopal-Krishna \etal 1995]{gop95} Gopal-Krishna, Giraud, E., Melnick, J., \& Della Valle, M. 1995, \aap, 303, 705 

\bibitem[Hales, Baldwin \& Warner 1993]{hal93} Hales, S., Baldwin, J., \& Warner, P. 1993, \mnras, 263, 25 

\bibitem[Hamann \& Ferland 1993]{ham93} Hamann, F., \& Ferland, G. 1993, \apj, 418, 11

\bibitem[Isobe, Feigelson \& Nelson 1986]{iso86} Isobe, T., Feigelson, E., \& Nelson, P. 1986, \apj, 306, 490 

\bibitem[Isobe \& Feigelson 1990]{iso90} Isobe, T. \& Feigelson, E. 1990, BAAS, 22, 917

\bibitem[Ivison \etal 1998]{ivi98} Ivison, R. \etal 1998, \mnras, 298, 583 

\bibitem[Iwamuro \etal 1996]{iwa96} Iwamuro, F., Oya, S., Tsukamoto, H., \& Maihara, T. 1996, \apjl, 466, L67 

\bibitem[Jarvis \etal 1999]{jar99} Jarvis, M., Rawlings, S., Willott, C., Blundell, K., Eales, S., \& Lacy, M. 1999, in Proc. Hy-fest, in press, astro-ph/9908106

\bibitem[Kapahi \etal 1998]{kap98} Kapahi, V., Athreya, R., van Breugel, W., McCarthy, P., \& Subrahmanya, C. 1998, \apjs, 118, 275 

\bibitem[Kurk \etal 1999]{kur99} Kurk, J., R\"ottgering, H., Pentericci, L., \& Miley, G. 1999, in ``Clustering At High Redshift'', ASP Conf. Series, ed. A. Mazuer \& O. Le~Fevre, in press, astro-ph/9910257

\bibitem[Lacy \etal 1994]{lac94} Lacy, M., \etal 1994, \mnras, 271, 504 

\bibitem[Lacy \etal 1999]{lac99} Lacy, M., Rawlings, S., Hill, G., Bunker, A., Ridgway, S., \& Stern, D. 1999, \mnras, 308, 1096 

\bibitem[Laing, Riley \& Longair 1983]{lai83} Laing, R., Riley, J., \& Longair, M. 1983, \mnras, 204, 151

\bibitem[Larkin \etal 2000]{lar00} Larkin, J., \etal, \apjl, in press, astro-ph/0002335

\bibitem[Lavalley, Isobe \& Feigelson 1992]{lav92} Lavalley, M., Isobe, T., \& Feigelson, E.  1992, ASP Conf. Ser. 25: Astronomical Data Analysis Software and Systems I, 1, 245 

\bibitem[Lawrence \etal 1986]{law86} Lawrence, C., Bennett, C., Hewitt, J., Langston, G., Klotz, S., Burke, B., \& Turner, K. 1986, \apjs, 61, 105 

\bibitem[Lilly \& Longair 1984]{lil84} Lilly, S., \& Longair, M. 1984, \mnras, 211, 833

\bibitem[Lilly 1988]{lil88} Lilly, S. 1988, \apj, 333, 161 

\bibitem[Lilly 1989]{lil89} Lilly, S. J. 1989, \apj, 340, 77

\bibitem[MacAlpine \etal 1985]{mac85} MacAlpine, G., Davidson, K., Gull, T., \& Wu, C. 1985, \apj, 294, 147 

\bibitem[Macklin 1982]{mac82} Macklin, J. 1982, \mnras, 199, 1119

\bibitem[McCarthy \etal 1987]{mcc87} McCarthy, P., van Breugel, W. , Spinrad, H., \& Djorgovski, S. 1987, \apjl, 321, L29 

\bibitem[McCarthy \etal 1990]{mcc90a} McCarthy, P., Kapahi, V., van Breugel, W., \& Subrahmanya, C. 1990, \aj, 100, 1014 

\bibitem[McCarthy \etal 1990]{mcc90b} McCarthy, P. \etal 1990, \apj, 365, 487 

\bibitem[McCarthy, van Breugel \& Kapahi 1991a]{mcc91a} McCarthy, P., van Breugel, W., \& Kapahi, V. K. 1991, \apj, 371, 478

\bibitem[McCarthy 1991b]{mcc91b} McCarthy, P. 1991, \aj, 102, 518 

\bibitem[McCarthy \etal 1991c]{mcc91c} McCarthy, P., van Breugel, W., Kapahi, V., \& Subrahmanya, C. 1991, \aj, 102, 522 

\bibitem[McCarthy 1993]{mcc93} McCarthy, P. 1993, \araa, 31, 639 

\bibitem[McCarthy \etal\ 1996]{mcc96} McCarthy, P., Kapahi, V., van Breugel, W., Persson, S., Atheya, R., \& Subrahmanya, C. 1996, \apjs, 107, 19

\bibitem[McCarthy \& Lawrence 2000]{mcc00} McCarthy, P. \& Lawrence, C. 2000, in preparation

\bibitem[Neeser \etal 1995]{nee95} Neeser, M., Eales, S., Law-Green, J., Leahy, J. \& Rawlings, S. 1995, \apj, 451, 76

\bibitem[Osterbrock 1989]{ost89} Osterbrock, D. E. 1989, Astrophysics of Gaseous Nebulae and Active Galactic Nuclei (Mill Valley, CA: Univ. Sci.)

\bibitem[Osterbrock \& Martel 1992]{ost92} Osterbrock, D. \& Martel, A. 1992, \pasp, 104, 76 

\bibitem[Pentericci \etal 2000]{pen00} Pentericci, L. \etal 1999, in preparation

\bibitem[Rawlings \& Saunders 1991]{raw91} Rawlings, S. \& Saunders, R. 1991, \nature, 349, 138 

\bibitem[Rawlings \etal 1996]{raw96} Rawlings, S., Lacy, M., Blundell, K., Eales, S, Bunker, A., \& Garrington, S. 1996, \nature, 383, 502

\bibitem[Rees 1990]{ree90} Rees, N. 1990, \mnras, 244, 233 

\bibitem[Rengelink \etal\ 1997]{ren97} Rengelink, R. \etal\ 1997, \aap, 124, 259

\bibitem[Riley, Waldram \& Riley 1999]{ril99} Riley, J., Waldram, E., \& Riley, J. 1999, \mnras, 306, 31 

\bibitem[R\"ottgering \etal\ 1994]{rot94} R\"ottgering, H., Lacy, M., Miley, G., Chambers, K., \& Saunders, R., \aasup, 108, 79

\bibitem[R\"ottgering \etal 1995]{rot95} R\"ottgering, H., Hunstead, R., Miley, G., van Ojik, R., \& Wieringa, M. 1995, \mnras, 277, 389

\bibitem[R\"ottgering \etal 1997]{rot97} R\"ottgering, H., van Ojik, R., Miley, G., Chambers, K., van Breugel, W., \& de Koff, S. 1997, \aap, 326, 505

\bibitem[Serjeant \etal 1998]{ser98} Serjeant, S., Rawlings, S., Lacy, M., McMahon, R., Lawrence, A., Rowan-Robinson, M., \& Mountain, M. 1998, \mnras, 298, 321 

\bibitem[Simpson \etal 1999]{sim99} Simpson, C. \etal 1999, \apj, 525, 659 

\bibitem[Spinrad \etal\ 1985]{spi85} Spinrad, H., Djorgovski, S., Marr, J., \& Aguilar, L. 1985, \pasp, 97, 932

\bibitem[Spinrad \etal 1995]{spi95} Spinrad, H., Dey, A., \& Graham, J. 1995, \apj, 438, 51

\bibitem[Stern \etal 1999]{ste99} Stern, D., Dey, A., Spinrad, H., Maxfield, L., Dickinson, M., Schlegel, D., \& Gonz\'alez, R. 1999, \aj, 117, 1122 (S99)

\bibitem[Tran \etal 1998]{tra98} Tran, H., Cohen, M., Ogle, P., Goodrich, R., \& di Serego Alighieri, S. 1998, \apj, 500, 660 

\bibitem[Tran \etal 2000]{tra00} Tran, H., Cohen, M., \& Villar-Mart\'\i n 2000, \aj, in press, astro-ph/0004383

\bibitem[van Breugel \etal 1985]{wvb85} van Breugel, W., Filippenko, A., Heckman, T., \& Miley, G. 1985, \apj, 293, 83 

\bibitem[van Breugel \etal 1998]{wvb98} van Breugel, W., Stanford, S. A., Spinrad, H., Stern, D., \& Graham, J. R. 1998, \apj, 502, 614

\bibitem[van Breugel \etal 1999]{wvb99} van Breugel, W., De Breuck, C., Stanford, S. A., Stern, D., R\"ottgering, H., \& Miley, G. 1999, \apj, 518, 61

\bibitem[van Ojik \etal 1994]{oji94} van Ojik, R., R\"ottgering, H., Miley, G., Bremer, M. N., Macchetto, F, \& Chambers, K. 1994, \aap, 289, 54

\bibitem[van Ojik 1995]{oji95} van Ojik, R. 1995, PhD thesis, Rijksuniversiteit Leiden

\bibitem[van Ojik \etal 1996]{oji96} van Ojik, R., R\"ottgering, H., Carilli, C., Miley, G., Bremer, M., \& Machetto, F. 1996, \aap, 313, 25

\bibitem[van Ojik \etal 1997]{oji97} van Ojik, R., R\"ottgering, H., Miley, G., \& Hunstead, R. W. 1997, \aap, 317, 358

\bibitem[Vernet \etal 1999]{ver99} Vernet, J., Fosbury, R., Villar-Martin, M., Cohen, M., di Serego Alighieri, S., \& Cimatti, A 1999, in ``The Hy redshift Universe'', eds. A. Bunker \& W. van Breugel, ASP Conf. Series, Vol. 193, p102

\bibitem[Villar-Mart\'\i n, Binette \& Fosbury 1996]{vil96} Villar-Mart\'\i n, M., Binette, L., \& Fosbury, R. 1996, \aap, 312, 751

\bibitem[Villar-Mart\'\i n, Tadhunter \& Clark 1997]{vil97} Villar-Mart\'\i n, M., Tadhunter, C., \& Clark, N. 1997, \aap, 323, 21 (VM97)

\bibitem[Villar-Mart\'\i n, Binette \& Fosbury 1999a]{vil99a} Villar-Mart\'\i n, M., Binette, L., \& Fosbury, R. 1999, \aap, 346, 7

\bibitem[Villar-Mart\'\i n \etal 1999b]{vil99b} Villar-Mart\'\i n, M., Tadhunter, C., Morganti, R., Axon, D., \& Koekemoer, A. 1999, \mnras, 307, 24 

\bibitem[Villar-Mart\'\i n \etal 1999c]{vil99c} Villar-Mart\'\i n, M., Fosbury, R., Binette, L., Tadhunter, C., \& Rocca-Volmerange, B. 1999, \aap, 351, 47 

\bibitem[Waddington \etal\ 1999]{wad99} Waddington, I., Windhorst, R., Cohen, S., Partridge, R., Spinrad, H., \& Stern, D. 1999, \apjl, in press, astro-ph/9910069

\bibitem[Willott \etal 1999]{wil99} Willott, C., Rawlings, S., Blundell, K., \& Lacy, M. 1999, \mnras, 309, 1017 

\end{thebibliography}
\end{document}